\documentclass[preprint]{aastex62}

\usepackage{multirow}

\graphicspath{{./}{figures/}}

\shorttitle{Emission Line Fluxes}
\shortauthors{Pharo et al.}

\begin{document}

\title{A Catalog of Emission-Line Galaxies from the Faint Infrared Grism Survey: Studying Environmental Influence on Star Formation}

\correspondingauthor{John Pharo}
\email{john.pharo@asu.edu}

\author{John Pharo}
\affiliation{School of Earth and Space Exploration, Arizona State University, Tempe, AZ 85287, USA}

\author{Sangeeta Malhotra}
\affiliation{NASAs Goddard Space Flight Center, Astrophysics Science Division, Code 660, Greenbelt, MD 20771 USA}
\affiliation{School of Earth and Space Exploration, Arizona State University, Tempe, AZ 85287, USA}

\author{James E. Rhoads}
\affiliation{NASAs Goddard Space Flight Center, Astrophysics Science Division, Code 660, Greenbelt, MD 20771 USA}
\affiliation{School of Earth and Space Exploration, Arizona State University, Tempe, AZ 85287, USA}

\author{Norbert Pirzkal}
\affiliation{Space Telescope Science Institute, Baltimore, MD 21218, USA}

\author{Steven L. Finkelstein}
\affiliation{Department of Astronomy, The University of Texas at Austin, Austin, TX 78712, USA}

\author{Russell Ryan}
\affiliation{Space Telescope Science Institute, Baltimore, MD 21218, USA}

\author{Andrea Cimatti}
\affiliation{Department of Physics and Astronomy (DIFA), University of Bologna, Via Gobetti 93/2, I-40129, Bologna, Italy}
\affiliation{INAF—Osservatorio Astrofisico di Arcetri, Largo E. Fermi 5, I-50125, Firenze, Italy}

\author{Lise Christensen}
\affiliation{1 Dark Cosmology Centre, Niels Bohr Institute, University of Copenhagen, Juliane Maries Vej 30, DK-2100 Copenhagen, Denmark}

\author{Nimish Hathi}
\affiliation{Space Telescope Science Institute, Baltimore, MD 21218, USA}

\author{Anton Koekemoer}
\affiliation{Space Telescope Science Institute, Baltimore, MD 21218, USA}

\author{Santosh Harish}
\affiliation{School of Earth and Space Exploration, Arizona State University, Tempe, AZ 85287, USA}

\author{Mark Smith}
\affiliation{School of Earth and Space Exploration, Arizona State University, Tempe, AZ 85287, USA}

\author{Amber Straughn}
\affiliation{NASAs Goddard Space Flight Center, Astrophysics Science Division, Code 660, Greenbelt, MD 20771 USA}

\author{Rogier Windhorst}
\affiliation{School of Earth and Space Exploration, Arizona State University, Tempe, AZ 85287, USA}

\author{Ignacio Ferreras}
\affiliation{Mullard Space Science Laboratory, University College London, Holmbury St. Mary, Dorking, Surrey RH5 6NT, UK}

\author{Caryl Gronwall}
\affiliation{Department of Astronomy \& Astrophysics, Pennsylvania State University, University Park, PA 16802}
\affiliation{Institute for Gravitation \& the Cosmos, Pennsylvania State University, University Park, PA 16802}

\author{Pascale Hibon}
\affiliation{ESO, Alonso de Cordova 3107, Santiago, Chile}

\author{Rebecca Larson}
\affiliation{Department of Astronomy, The University of Texas at Austin, Austin, TX 78712, USA}

\author{Robert O'Connell}
\affiliation{The University of Virginia, Charlottesville, VA 22904-4325, USA}

\author{Anna Pasquali}
\affiliation{Astronomisches Rechen-Institut, Zentrum fuer Astronomie, Universitaet Heidelberg, Moenchhofstrasse 12-14, D-69120 Heidelberg, Germany}

\author{Vithal Tilvi}
\affiliation{School of Earth and Space Exploration, Arizona State University, Tempe, AZ 85287, USA}

\begin{abstract}

We present a catalog of 208 $0.3 < z < 2.1$ Emission Line Galaxies (ELG) selected from 1D slitless spectroscopy obtained using Hubble's WFC3 G102 grism, as part of the Faint Infrared Grism Survey (FIGS). We identify ELG candidates by searching for significant peaks in all continuum-subtracted G102 spectra, and, where possible, confirm candidates by identifying consistent emission lines in other available spectra or with published spectroscopic redshifts. We provide derived emission line fluxes and errors, redshifts, and equivalent widths (EW) for H$\alpha$ $\lambda6563$, [O\textsc{iii}]$\lambda\lambda4959,5007$, and [O\textsc{ii}]$\lambda\lambda3727$ emission lines, for emission line galaxies down to AB(F105W) $ > 28$ and $> 10^{-17}$ erg cm$^{-2}$ s$^{-1}$ line flux. We use the resulting line catalog to investigate a possible relationship between line emission and a galaxy's environment. We use 7th-nearest-neighbor distances to investigate the typical surroundings of ELGs compared to non-ELGs, and we find that [O\textsc{iii}] emitters are preferentially found at intermediate galaxy densities near galaxy groups. We characterize these ELGs in terms of the galaxy specific star formation rate (SSFR) versus stellar mass, and find no significant influence of environment on that relation. We calculate star formation rates (SFR), and find no dependence of SFR on local galaxy surface density for $0.3 < z < 0.8$ H$\alpha$ emitters and for $0.8<z<1.3$ [O\textsc{iii}] emitters. We find similar rates of close-pair interaction between ELGs and non-ELGs. For galaxy surface densities $\Sigma \leq 30$ Mpc$^{-2}$, we find no consistent effect of environment on star formation.

\end{abstract}



\section{Introduction} \label{sec:intro}

The detection and measurement of nebular line emission in galaxies has long been a useful tool in the study of galaxy evolution. Hydrogen recombination lines, such as H$\alpha$ $\lambda$6563 and H$\beta$ $\lambda$4861, and forbidden transitions in ionized oxygen, such as [O\textsc{iii}]$\lambda\lambda4959,5007$ and [O\textsc{ii}]$\lambda 3727$, can be used to measure a galaxy's recent star formation \citep{ken98}, its gas-phase composition \citep{kz99}, and dust extinction \citep{cal94}, among other properties. Furthermore, the spectroscopic identification of an emission line enables the measurement of a galaxy's redshift with much higher precision than is achievable with methods relying on broadband photometry alone, even with relatively low-resolution slitless spectroscopy or narrowband photometry \citep{xu07, xia11, fer14, pha18}. Measurements of fundamental galaxy properties like luminosity, as well as assessments of a galaxy's interactions with nearby galaxies and their environments, rely in part on this measure of distance.


A key aspect of the study of galaxy evolution is the potential influence of a galaxy's surrounding environment on its development, particularly in how it relates to star formation. In the local universe, red, passive galaxies are associated with overdensities, while blue galaxies with active star formation (of which line emission is an indicator) are more likely found in low-density environments and lower-mass dark matter halos \citep{dre80, bal04, kau04} and in galaxies with lower stellar mass \citep{pas09}. At higher redshift, the picture is less clear, with some studies matching the local result \citep{pat09}, while others find a weak star formation dependence on environment \citep{gru11, sco13}. \citet{elb07} and \citet{coo08} find that star formation activity correlates with density at high redshift, and \citet{tra10} find a high level of star formation in a cluster core at $z=1.62$ compared to lower densities, the opposite of the local trend. \citet{sob11} and \citet{dar14} report an increase in star formation at intermediate density, potentially associated with groups or filaments rather than rich clusters. In order to make clearer sense of this picture, further studies capable of accurately measuring both local environments and star formation are needed. The identification of line emission in galaxies can achieve this purpose.

Emission line galaxies (ELGs) can be detected in several ways. In principle, the most straightforward method is the use of ground-based spectroscopy, but this is not always practical for the faintest objects, requiring some pre-selection of targets and spectral features. Another common approach is the use of narrowband photometric surveys \citep[e.g.,][]{bor93, rho01, gea08, sob11, sob12, cou18}, which detect emission lines by measuring the flux excess between narrowband photometry and nearby broadband photometry. This method is useful for obtaining a large number of detections, but these detections are limited to a narrow redshift window defined by the width of the narrowband.

A third approach for ELG detection is the use of low-resolution, slitless spectroscopy. Recent surveys have made use of the Hubble Space Telescope's (HST) ACS (APPLES, \citeauthor{pas03}, \citeyear{pas03}; GRAPES, \citeauthor{pir04, mal05, rho05}, \citeyear{pir04}; PEARS, \citeauthor{pir09, rho13}, \citeyear{pir09}), Wide-Field Camera 3 (WFC3) G102 (WISP, \citeauthor{ate10}, \citeyear{ate10}; GLASS, \citeauthor{tre15}, \citeyear{tre15}; FIGS, \citeauthor{pir18}, \citeyear{pir18}), and G141 (WISP, \citeauthor{ate10}, \citeyear{ate10}; GLASS, \citeauthor{tre15}, \citeyear{tre15}; 3D-HST, \citeauthor{mom16}, \citeyear{mom16}) grisms to identify ELGs over a broad redshift range ($0 < z < 7.5$) and without a pre-selection of targets that might exclude continuum-faint sources \citep{rho13, til16, lar18}. In the Faint Infrared Grism Survey \citep[FIGS;][]{pir17}, we apply this approach with deep WFC3 G102 observations taken at multiple position angles in order to maximize emission line sensitivity, minimize spectral contamination, and more accurately measure the central wavelengths of emission lines \citep{xu07, str08, str09, xia11, xia12, pir13}.

The FIGS grism data therefore provides an opportunity to study how an ELG's emission properties relate to its environment. First, the slitless grism selection enables the unbiased detection of continuum-faint ELGs, which can be used for a study of star formation in the FIGS fields. Second, grism studies have shown that the combination of grism spectroscopy with broadband photometry can significantly improve photometric redshift accuracy \citep{rya07, bra12, pha18}, and that these improved redshift catalogs can be used to better identify and study galaxy overdensities \citep{pha18}. With FIGS data, we can then measure emission lines and star formation rates (SFRs) across a broad redshift range, and evaluate their local environments using improved grism redshifts.

In this paper, we present a catalog of emission line galaxies derived from an automated search of 1D slitless spectra from FIGS obtained with HST's WFC3 G102 grism. In this catalog, we list the line fluxes, redshifts, and observed equivalent widths (EW) for 208 strong-line (H$\alpha$, [O\textsc{iii}]$\lambda\lambda5007,4959$, and [O\textsc{ii}]$\lambda 3727$) emitters in a redshift range of $0.3 < z < 2.1$. We then combine this catalog with a previous study of overdensities in the FIGS fields \citep{pha18} to study ELG properties as a function of their local galaxy environment. In \S2, we briefly describe the FIGS data collection and reduction, as well as the sources of additional spectra we used to supplement our study. In \S3, we detail our search methods for identifying and confirming ELG candidates in 1D spectra. In \S4, we describe our line flux measurements, present the final ELG catalog, and summarize its properties. In \S5, we study these properties as functions of the local environment and stellar mass. For this paper we will use $H_0 = 67.3$ km s$^{-1}$ Mpc$^{-1}$ and $\Omega_M = 0.315$, $\Omega_{\Lambda} = 0.685$ \citep{pla15}. All magnitudes are given in the AB system \citep{og83}.

\section{Survey Description and Data} \label{sec:survey}

\subsection{FIGS Observations}

The Faint Infrared Grism Survey (FIGS, HST Cycle 22, PID:13779, PI S. Malhotra) used the HST WFC3-G102 infrared grism (see Figure \ref{fig:spec}) to obtain deep slitless spectroscopy of $\sim$ 6000 galaxies. FIGS achieved 40-orbit depth in 4 fields within the greater GOODS-North and GOODS-South fields \citep{gia04}, designated GN1, GN2, GS1 (HUDF), and GS2 (HDF-PAR2). Objects in each field were observed in 5 different 8-orbit position angles (PAs) in order to mitigate the contamination of spectra from overlapping spectra from nearby objects. The grism image at each PA covers a 2.05'x2.27' field of view, and the G102 grism has a resolution of 24.5 \AA\ per pixel. The total area of coverage over all fields is 17.7 square arcminutes. 

\subsection{Spectral Extraction and Properties}

In this paper, we used 1D spectra of individual PAs which were generated using the methods described in \citet{pir17}. Here we briefly summarize this process. We reduced FIGS data  in a manner that loosely follows the method used for GRAPES and PEARS, the previous deep HST grism surveys \citep{pir04, mal05, xu07, rho09, str09, xia12, pir13}. First, we generated a master catalog of sources from deep CANDELS survey mosaics in the F850LP filter in ACS and the F125W and F160W filters in WFC3 \citep[approximately the z, J, and H bands;][]{gro11, koe11}. We astrometrically corrected the data to match the absolute astrometry of the GOODS V2.0 catalogs. We estimated the background levels of the grism observations by using a two-component model, which included a constant Zodiacal light background, as well as a varying He \textsc{i} light background. To generate the individual spectra, we used a Simulation Based Extraction (SBE) approach that accounts for spectral contamination from overlapping spectra, as well as allowing the use of an optimal extraction approach \citep{hor86} when generating 1D spectra from 2D spectra. We refer the reader to \citet{pir17} for a complete description of these processes. \citet{pir17} also describes a method for combining individual PAs into a single 1D spectrum, though we do not use the combined spectra here. Individual PAs are subject to different levels of contamination, and depending on the location of the emission line region, they may not all exhibit the same line emission. For these reasons, we use the individual PA spectra for our ELG search and analysis. 

\begin{figure}
\plotone{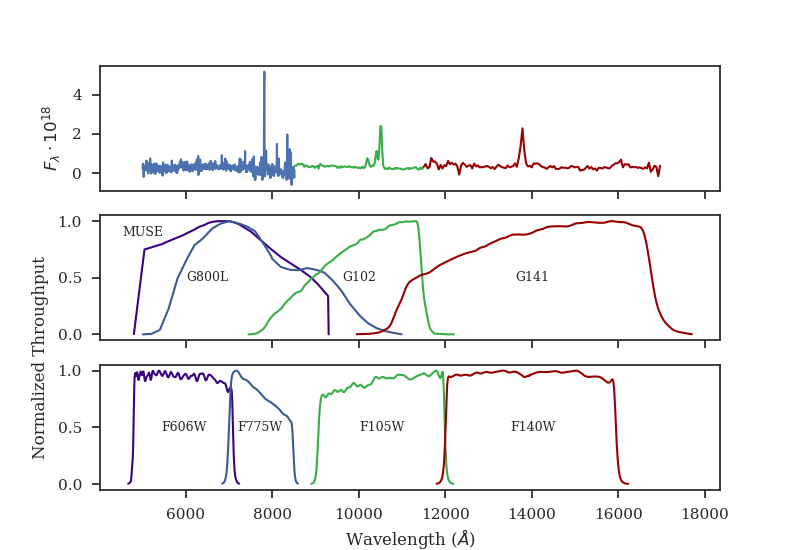}
\caption{The top panel shows the spectrum of an example ELG from FIGS at z=1.098, with spectra from MUSE/VLT (dark blue), the HST ACS G800L grism (blue), the HST WFC3 G102 grism (green), and the G141 grism (red). The y-axis for this panel is in units of $10^{-18}$ ergs s$^{-1}$ cm$^{-2} \AA^{-1}$. The middle panel gives the throughput curves for each spectrograph, normalized to the maximum throughput of each. The lower panel shows the throughput curves for HST photometric bands at comparable wavelengths. In this example, [O\textsc{iii}]$\lambda\lambda$4959,5007 are detected in the FIGS G102 spectrum, with H$\alpha$+[N\textsc{ii}] detected in G141, and [O\textsc{ii}]3727 detected in MUSE/VLT.\label{fig:spec}}
\end{figure}

We initially extracted all sources down to a continuum level of F105W $< 30$ mag. When the extractions were complete, we had an average of $\sim1700$ spectra per field, with a typical $3\sigma$ continuum detection limit of $m_{F105W} = 26$ mag and an emission line sensitivity of $10^{-17}$ erg cm$^{-2}$ s$^{-1}$. The middle panel of Figure \ref{fig:spec} shows the throughput curve for the G102 grism compared to other spectral and broadband curves. We restricted use of the G102 spectra to wavelengths between 8500 and 11500 \AA, where the grism throughput is greater than 20\%.

\subsection{Additional Spectral Data}

\subsubsection{MUSE/VLT}

 For the GS1/HUDF FIGS field, we supplemented our infrared FIGS spectra with deep archival high-resolution optical IFU spectra taken with the Multi-Unit Spectroscopic Explorer (MUSE) instrument \citep{bac10} from the Very Large Telescope (VLT). This expands the available spectroscopic wavelength coverage for the GS1 field considerably, enabling confirmation of detected emission lines in FIGS via the identification of complementary emission lines at optical wavelengths, even for many faint sources. We used the publicly available IFU spectra from the MUSE Hubble Ultra Deep Survey \citep{bac17}, a mosaic of nine $1 \times 1$ arcmin$^2$ MUSE fields in the HUDF. The data were reduced using standard procedures and ESO pipelines. After the initial reduction steps of each observations, offsets between pointings in the mosaic were found by cross correlations and matching stars at the areas of overlapping mosaics. The 180 cubes were then combined with ESOrex again.  This product was cleaned of strong sky subtraction residuals using ZAP (Zurich Atmosphere Purging) by \citet{sot16}.
 
 In order to extract spectra for emission-line objects in our sample, we first used the known sky coordinates for each object in FIGS to find RA-Dec matches in the MUSE datacube. At each wavelength slice, we placed a 2\arcsec\ aperture centered on the object, which we determined was able to capture the total flux from most line-emitting sources at the redshifts considered. Then, we generated 1D spectra for the matched objects by summing up the flux within the aperture at each wavelength, across the entire MUSE wavelength range (see Figure \ref{fig:spec} for the MUSE wavelength coverage compared to WFC3 G102), using the MPDAF software package \citep{bac16}. This produced a catalog of extracted FIGS candidate spectra from the reduced MUSE datacube. The MUSE data wavelength coverage extends from 4752 \AA\ to 9347 \AA\ with a spectral resolution of 2.3 \AA, though the sensitivity begins to drop off at wavelengths lower than 5000 \AA, and at wavelengths higher than 9200 \AA, the noise from sky emission begins to dominate, so we restrict our usage of the MUSE spectra to between these wavelengths. MUSE has a $3\sigma$ line sensitivity of $\sim 3 \cdot 10^{-19}$ erg cm$^{-2}$ s$^{-1}$, and thus should detect lines of strength comparable to or even considerably less than the lines found in FIGS G102 spectra.
 
\newpage

\subsubsection{G800 Grism Data}

In the GS1/HUDF field, we were also able to make use of line identifications from the Grism ACS Program for Extragalactic Science (GRAPES) \citep{xu07}, which used the G800L grism from HST ACS, a low-resolution (40 \AA\ per pixel at 8000 \AA) optical grism. \citet{xu07} were able to identify lines from $\sim6000$ \AA\ to $\sim9500$ \AA, a similar region of coverage to VLT/MUSE. This enabled us to search for complementary optical lines for FIGS sources while simultaneously confirming some ELGs from GRAPES.  

\subsubsection{G141 Grism Data}

In the other FIGS fields (GN1, GN2, GS2), we also made use of archival WFC3 G141 grism spectra \citep[HST proposal ID 13266]{rya13} collected from the WISP \citep{ate10} and 3D-HST \citep{bra12, mom16} surveys. Inclusion of this data effectively extended the FIGS spectral coverage out to $\sim1.7$ micron. These additional collected G141 spectra are not as deep as the FIGS G102 data, with $>90\%$ completeness down to J$<24$ mag. They also have lower spectral resolution than the WFC3 G102 spectra (46.5 \AA\ per pixel at 14000 \AA) and thus were of limited use for candidate confirmation, but they allowed for the detection of strong line emission in some objects.

\subsubsection{Spectroscopic Redshifts}

We assembled compilations of published high-quality spectroscopic redshifts (spec-zs) in the GOODS-N and CDFS fields (N. Hathi, private communication). These fields are well-studied, and the existence of independent spectroscopy allowed us to confirm the emission-line-derived redshifts (and thus, the identified emission line) of some of our brighter sources. Many of the published spec-z catalogs included quality designations distinguishing the reliability of different spectra. The exact scales of quality used differed somewhat between surveys, but we generally used only those results deemed ``good" or better by the original survey. Our compilations included spec-zs from \citet{wir04}, \citet{mal05}, \citet{gra06}, \citet{pas06}, \citet{red06}, \citet{rav07}, \citet{bar08}, \citet{hat08}, \citet{str08}, \citet{van08}, \citet{wuy08}, \citet{fer09}, \citet{hat09}, \citet{rho09}, \citet{str09}, \citet{van09}, \citet{wuy09}, \citet{bal10}, \citet{sil10}, \citet{yos10}, \citet{coo11}, \citet{xue11}, \citet{coo12}, \citet{ono12}, \citet{fin13}, \citet{kur13}, \citet{lef13}, \citet{pir13}, \citet{tru13}, \citet{son14}, \citet{kri15}, \citet{lef15}, \citet{mor15}, \citet{wir15}, \citet{tru15}, \citet{mom16}, \citet{her17}, \citet{ina17}, and \citet{mcl18}.

\section{Emission Line Identification Methods} \label{sec:id}

\subsection{Search for ELG Candidates}

We conducted a blind search for ELGs among the FIGS 1D spectra. Because we obtained our infrared spectra via slitless grism spectroscopy, there was no pre-selection of ELG candidates before the search via the placement of slits or by broadband magnitude cutoffs beyond the survey depth. This had the advantage of enabling the detection of ELGs with potentially very low continuum levels, and so might allow for the identification and study of smaller and/or fainter galaxies with nebular line emission. However, this did require an efficient method for selecting ELG candidates from the total sample of FIGS spectra. In order to search the $\sim 6000$ FIGS spectra for emission lines, we developed a code to automatically search for and identify significant peaks in a 1D spectrum. 

First, the level of the continuum flux had to be estimated at each wavelength element in the 1D spectrum. To measure this, we used a median-flux filter, which assumes a prospective line width and calculates the local continuum from the median flux outside that line width, in wavelength regions on either side of the line. A given wavelength $\lambda_0$ is taken to be the center of a potential line. The potential line flux is measured as all the flux contained within a line width $2\Delta \lambda_1$, so that the algorithm defines the potential line as the region covered by:

\begin{equation}
\lambda_0 - \Delta \lambda_1 < \lambda < \lambda_0 + \Delta \lambda_1
\end{equation}

\noindent Then the code estimates the nearby continuum by looking at regions to either side of the line with width $\Delta \lambda_2$. The nearby continuum is defined then as the regions contained in: 

\begin{equation}
(\lambda_0 - \Delta \lambda_1) - \Delta \lambda_2 < \lambda < (\lambda_0 - \Delta \lambda_1) \text{  and  }  (\lambda_0 + \Delta \lambda_1) < \lambda < (\lambda + \Delta \lambda_1) + \Delta \lambda_2
\end{equation}

\begin{figure}
\plotone{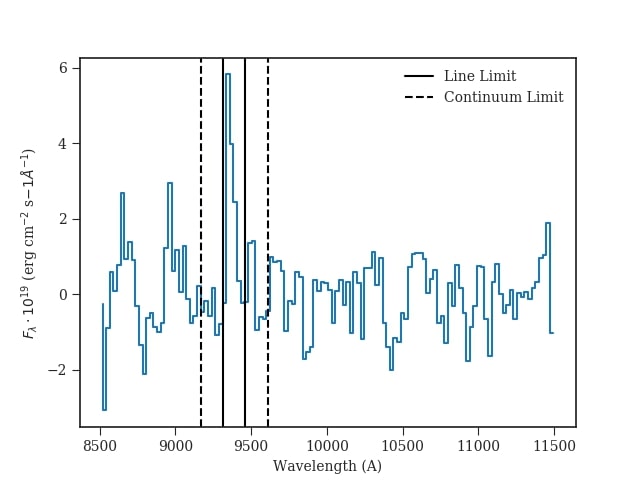}
\caption{The continuum-subtracted 1D spectrum of one position angle of one FIGS ELG: ID GS1-2375. This figure shows a snapshot of the continuum-subtraction and line-finding routine. The routine identifies a ``test line" region with a given pixel width, shown here as the region contained within the solid vertical lines. Next the routine uses the pixels between the solid lines and the dashed lines to estimate the local continuum flux, and subtracts that flux from the test line. Then the routine estimates the S/N ratio of the continuum-subtracted test line, and if the ratio surpasses the $5\sigma$ threshold, it reports a possible detection. This process iterates over each wavelength element in the spectrum.\label{fig:line_ex}}
\end{figure}

\noindent The algorithm then takes the median flux of the wavelength pixels constrained by Equation 2 as an estimate of the local continuum around the hypothetical line, and subtracts this flux from the flux at $\lambda_0$ in order to obtain the continuum-subtracted or residual flux at that point. If there is a line present at $\lambda_0$, this method allows for measurement of the level of the continuum without influence from the line flux. The code takes the standard deviation among this set of continuum fluxes as an estimate of the flux error at $\lambda_0$. If $\lambda_0$ is too near the edge of the spectrum to measure $\Delta \lambda_2$ on both sides, we estimate the continuum based on the fluxes from just the complete side. The algorithm repeats this process, iterating over each wavelength element in a given spectrum, estimating the continuum flux at that wavelength, and subtracting it. We refer to Figure \ref{fig:line_ex} for an example continuum-subtracted spectrum. We were able to best minimize false detections while retaining real ones with $2\Delta \lambda_1 = 122.5$ \AA\ and $\Delta \lambda_2 = 147$ \AA, based on tests of variable $\Delta \lambda_1$ and $\Delta \lambda_2$ with a preliminary subsample of ELGs with matching spectroscopic redshifts.


After the spectrum is continuum-subtracted, the code calculates the signal-to-noise ratio ($S/N$) at each wavelength with the residual flux and the flux error, once more iterating through the list of wavelength elements. The sum of the fluxes constrained by Equation 1 determines the hypothetical line signal, and the estimated flux errors added in quadrature measure the noise of the hypothetical line. After this calculation is complete for all wavelengths, the algorithm identifies the location with maximum line $S/N$ in the spectrum. If $S/N > 5$, we fit a Gaussian at the central wavelength element from which we obtain a measure of the continuum-subtracted integrated line flux. The code then subtracts the fit line from the residual flux spectrum and checks the next-highest $S/N$. If the $S/N$ ratio still exceeds 5, the routine repeats until the peak $S/N$ ratio is below the detection threshold.


We run this routine on the individual PA spectra in each field, and record all instances of $S/N > 5$. If the code finds a peak in at least two PAs with centroids at the same or adjacent wavelength elements (24.5 \AA\ in either direction), it declares a detection. Lower $S/N$ thresholds produced numerous false positives, so we used the $S/N > 5$ cutoff to maintain a more robust sample. We avoided using simultaneous fits of all PAs in order to avoid including contaminated PAs in a combined significance measurement. With individual PA fits, contaminated detections could more easily be identified and removed. After running the routine over all galaxies in the field, the list of detections forms an ELG candidate list.

\subsection{Line Identification and Confirmation}

Once we had obtained lists of candidate detections for each field, we next attempted to identify the type of emission line detected in each spectrum. First, we matched the candidate lists to our photometric redshift (photo-z) catalog \citep{pha18} and assigned a preliminary line ID based on the likely redshift. For the purposes of this result, we focused on three strong line IDs that could be robustly detected at FIGS resolution and sensitivity: H$\alpha \lambda 6563$, [O\textsc{iii}]$\lambda \lambda$4959,5007, and [O\textsc{ii}]$\lambda$3727. We did this because these lines typically have the strongest emission, and therefore can be detected robustly, and because they are common features of star-forming galaxies. H$\beta$4861 could theoretically be resolved and detected alongside [O\textsc{iii}], but was typically faint enough that it was difficult to detect at a significant level. Other FIGS studies have looked at Ly$\alpha$ line emission at higher redshifts \citep{til16, lar18}. We measure detections of GS2-1406 at levels of significance $\geq3.5\sigma$, consistent with those measured by \citet{lar18}, though these are lower than the $5\sigma$ cut used in this work. We detect GN1-1292 significantly in only one of the two PAs where it is measured in \citet{til16}. \citet{pir18} also detects GS2-1406 but not GN1-1292.


After the preliminary photo-z identification, we sought to confirm the existence and type of the line by checking the detection against ancillary data. The most straightforward way to do this was to check for other emission lines. Since the wavelength ratio between a given pair of emission lines is invariant across redshift, the detection of two strong lines is a useful check. For eight candidates, two strong lines were measured in the FIGS G102 spectra alone, and for 59 others we identified pairs by checking matched ACS/G800L, MUSE/VLT, and WFC/G141 spectra (described in \S2.3). This most commonly involved finding H$\alpha$-[O\textsc{iii}] and [O\textsc{iii}]-[O\textsc{ii}] pairs. Occasionally, we were able to identify another spectral feature, such as a strong 4000 \AA\ break, in order to confirm the redshift. We note that while finding a matching line can confirm a line detection, not finding a matching line does not necessarily mean the detection is false, since the true relative line strengths are not known ahead of time. The matching line may be sufficiently weaker than the FIGS line, or the matching spectra sufficiently shallow, such that the matching line is not detected. 

If matching spectra were not available, or a strong line was not identified, we next checked for a matching spectroscopic redshift (spec-z). If a spec-z assigned the line peak-wavelength a restframe wavelength that matched an emission line within the wavelength range of a FIGS grism element, we assigned the line the spec-z ID. If neither matching lines nor spec-z IDs were available, then we let the photo-z identification stand. 

In each field, there were a handful of objects with a significant detection but no good redshift measurement. These were almost all very continuum faint (F105W $> 27.5$ mag) objects, which both reduced the availability of broadband fluxes to use for photometric redshift fits and made the spectra more susceptible to contamination from nearby sources. Consequently, most of these candidates were removed through visual inspection, leaving four likely ELGs with no redshift: GS1-1062, GS2-532, GS2-838, and GS2-1624.

With the lines identified, we compared our results to the line list derived from \citet{pir18}, a study of FIGS strong line emitters using a distinct identification method with FIGS 2D spectra. In this paper, we do an independent selection and measurement of ELGs so as not to bias the findings of different search methods. However, we have compared our line candidates with those found in \citet{pir18} and find them to be in close agreement, with a 90\% match in identifications. A complete match would have been highly unlikely, as the methods have different strengths. The 2D method performs better at identifying broader lines that are wider than the median filter used with the 1D search. As described in \S3.1, we experimented with different median filter widths when designing the 1D search, and found that expanding the filter too broadly tended to introduce false detections and make it more difficult to detect fainter objects, since the detection becomes more susceptible to changes in the continuum. The 2D method, however, requires detections in at least 3 PAs. For a given FIGS field, the different roll angles do not overlap completely, so near the edges of the field there are regions without the full 5 PA coverage. In these regions, it is easier to get a detection with the 1D method, as it requires detections in only 2 PAs.


\newpage

\section{Flux Catalog} \label{sec:flux}

\begin{deluxetable*}{ccccccc}
\tablecaption{Median Properties of Emission Line Galaxies\label{tab:elgs}}
\tablecolumns{6}
\tablehead{
\colhead{Line} &
\colhead{Number} &
\colhead{z} &
\colhead{Flux (erg cm$^{-2}$ s$^{-1}$)} &
\colhead{$\sigma_F / F$\tablenotemark{a}} &
\colhead{F105W Mag} &
\colhead{EW (\AA)}
}
\startdata
H$\alpha \lambda 6563$ & 71 & 0.56 & $8.1\cdot10^{-17}$ & 14\% & 22.9 & 42 \\\relax
[O\textsc{iii}] $\lambda\lambda$4959,5007 & 81 & 0.99 & $5.3\cdot10^{-17}$ & 21\% & 24.3 & 67 \\\relax
[O\textsc{ii}] $\lambda$3727 & 56 & 1.76 & $3.5\cdot10^{-17}$ & 20\% & 24.2 & 53 \\
\enddata
\tablenotetext{a}{The median flux error as a percentage of the median line flux.}
\end{deluxetable*}

With a robust ELG list, we next systematically fit the strong emission lines in order to obtain flux measurements and more precise line centers (and therefore redshifts). To do this, we used a combined Gaussian fit to the line and power-law fit to the local continuum, using a Python coding package called \textit{lmfit} \citep{new14}. The peak-finding routine (\S3.1) provided first estimates for the wavelength of the line center and the flux level of the nearby continuum. We restricted the possible wavelength of the line center allowed by the fit to only vary by the width of one grism element in either direction from this initial guess. 

For the H$\alpha$ line, the nearby [N\textsc{ii}]$\lambda\lambda$6548,6584 lines are blended with H$\alpha$ in the G102 grism, so that our recorded H$\alpha$ fluxes are actually the combined fluxes of these three lines. \citet{fai18} have derived an empirical estimate of the [N\textsc{ii}]/H$\alpha$ ratio in G102 as a function of redshift and stellar mass for $0 < z < 2.7$ and $8.5 < \log(M_{\star}/M_{\odot}) < 11.0$. This empirical relation gives a fractional flux ranging from 5\% to 45\%. For the [O\textsc{iii}]$\lambda\lambda$4959,5007 lines, we simultaneously fit two Gaussians and the continuum, with an additional restriction that the flux ratio of the two Gaussians match the theoretical intensity ratio of 2.98 for the two lines derived in \citet{sz00}. The [O\textsc{ii}]$\lambda\lambda 3727,3729$ doublet is too close to be resolved separately in FIGS spectra, and so is measured and reported together.

We ran this fitting procedure on each strong line in each PA that yielded a detection, and we averaged the fits for each line to obtain a final observed flux measurement. To get the flux error, we first estimated the error of the flux of each pixel from the standard deviation in the flux of the nearby continuum pixels. Then we integrated these errors with the line fit to produce the total error for the integrated line flux, rather than simply use the error in the fit parameters. This method typically produced a larger and more realistic flux error than relying on the derived error of the fit parameters, which was often artificially small resulting from the constraints on the fit.

\begin{figure}
\plotone{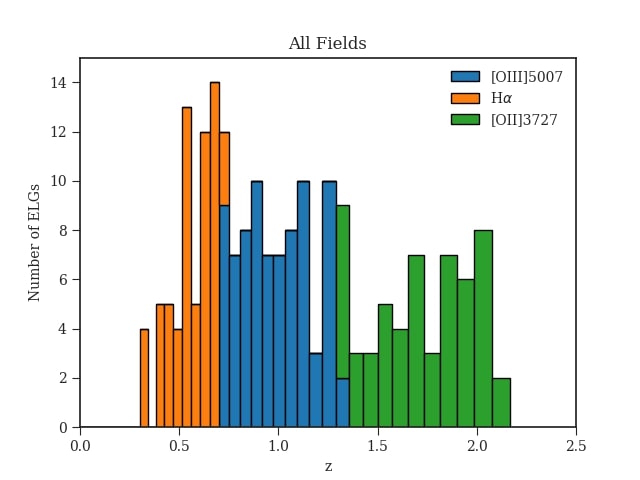}
\caption{The distribution of identified ELGs by redshift. The histogram bins are scaled by $\Delta z = 0.03 \cdot (1 + z)$ in order to encompass the expected redshift error derived from our redshift catalogs (though many individual objects have additional spectroscopic confirmation, and thus their real error is much lower). The bars of the histogram are colored according to the FIGS strong line ID, and redshift bins that contain more than one type of line in FIGS have stacked bars of two colors, so that the height of the stack is still the total number of objects in the bin. \label{fig:z_hist}}
\end{figure}

\begin{figure}
\plotone{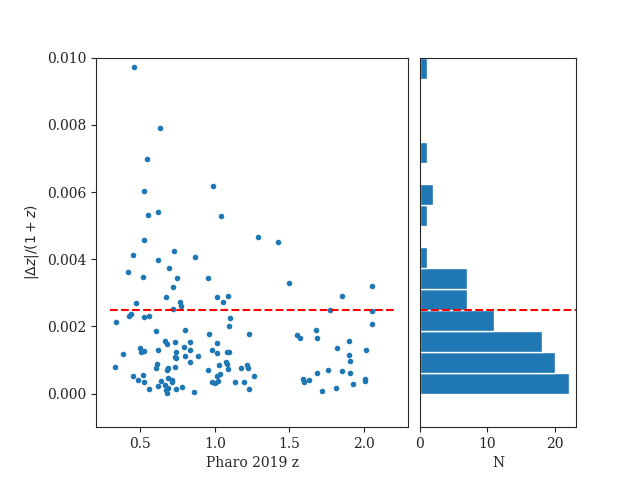}
\caption{A comparison of line-derived redshifts between the 1D sample and matching 2D-selected ELGs from \citet{pir18}. The dashed line shows $|\Delta z|/(1+z) = 0.0025$. These offsets are within the scatter observed by \citet{pir18} comparing derived redshifts between individual PAs, and thus could be explained by wavelength offsets between different line regions. \label{fig:z_comp}}
\end{figure}

We summarize the median properties of each type of strong line-emitter in Table 1. We give the full emission line catalog, including individual line fluxes, redshifts, continuum magnitudes, and equivalents widths in Table 2 in Appendix A. Figure \ref{fig:z_hist} gives the redshift distribution of the lines, covering $0.3 < z < 2.1$, which is the full redshift coverage sampled by these three strong lines. Each line type's redshift distribution is set by the wavelength coverage of the grism, though there is some overlap between H$\alpha$ and [O\textsc{iii}] and between [O\textsc{iii}] and [O\textsc{ii}], as shown by the stacked bars. The bin sizes in the histogram scale with $0.03 \cdot (1+z)$, so that the bin sizes roughly correspond to the photometric redshift error derived in \citet{pha18}, the redshift binning used in the enrionment analysis in \S5. Comparing our redshift measurements to those in \citet{pir18}, we find close agreement, with a median absolute difference of $|z_1 - z_2| = 0.001 \cdot (1 + z_1)$. Figure \ref{fig:z_comp} shows a comparison as a function of redshift. We find 81 [O\textsc{iii}] emitters, more than each of the other two (71 H$\alpha$, 56 [O\textsc{ii}]), likely because it spans more volume coverage than the lower-z H$\alpha$ while having less redshift dimming than the higher-z [O\textsc{ii}].

\begin{figure}
\plotone{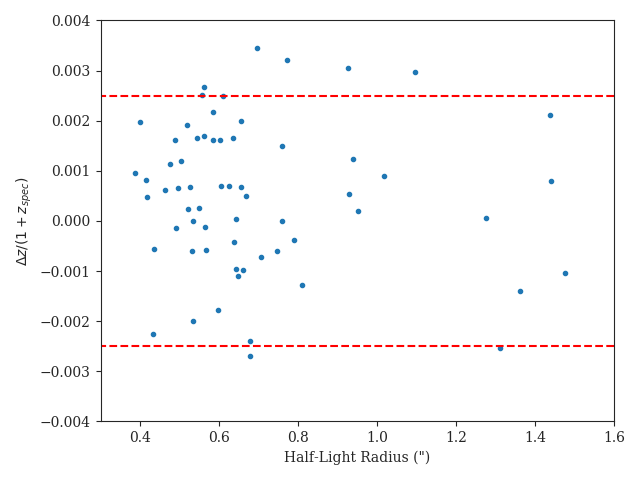}
\caption{A comparison of the line-derived redshifts from this study and the matching spectroscopic redshift sample from \S2.3.4 (excluding grism surveys) as a function of the half-light radius of the ELGs. The dashed lines show $\Delta z /(1+z) = \pm 0.0025$, the redshift change corresponding to one WFC3/G102 grism element at 10000 \AA. Two outliers with $\Delta z/(1+z) = -.065$ and $-.08$ are not shown. \label{fig:z_comp2}}
\end{figure}

We also compared the line-derived redshifts with the matching sample of high-quality spec-zs described in \S2.3.4, excluding the slitless grism surveys. We calculated $\Delta z /(1+z) = (z_{FIGS} - z_{spec}) / (1+z_{spec})$. Figure \ref{fig:z_comp2} shows $\Delta z /(1+z)$ as a function of the galaxy half-light radius, taken from the catalogs in \citet{sk14}. There is no apparent dependence of the redshift accuracy on the size of the galaxy, and we find that 80\% of the matched ELGs have $|\Delta z| /(1+z) \leq 0.0025$, the redshift change corresponding to one WFC3/G102 pixel at 10000 \AA. 97\% of the matched ELGs have $|\Delta z| /(1+z) \leq 0.005$, with just two outliers.

\begin{figure}
    \includegraphics[width=\textwidth]{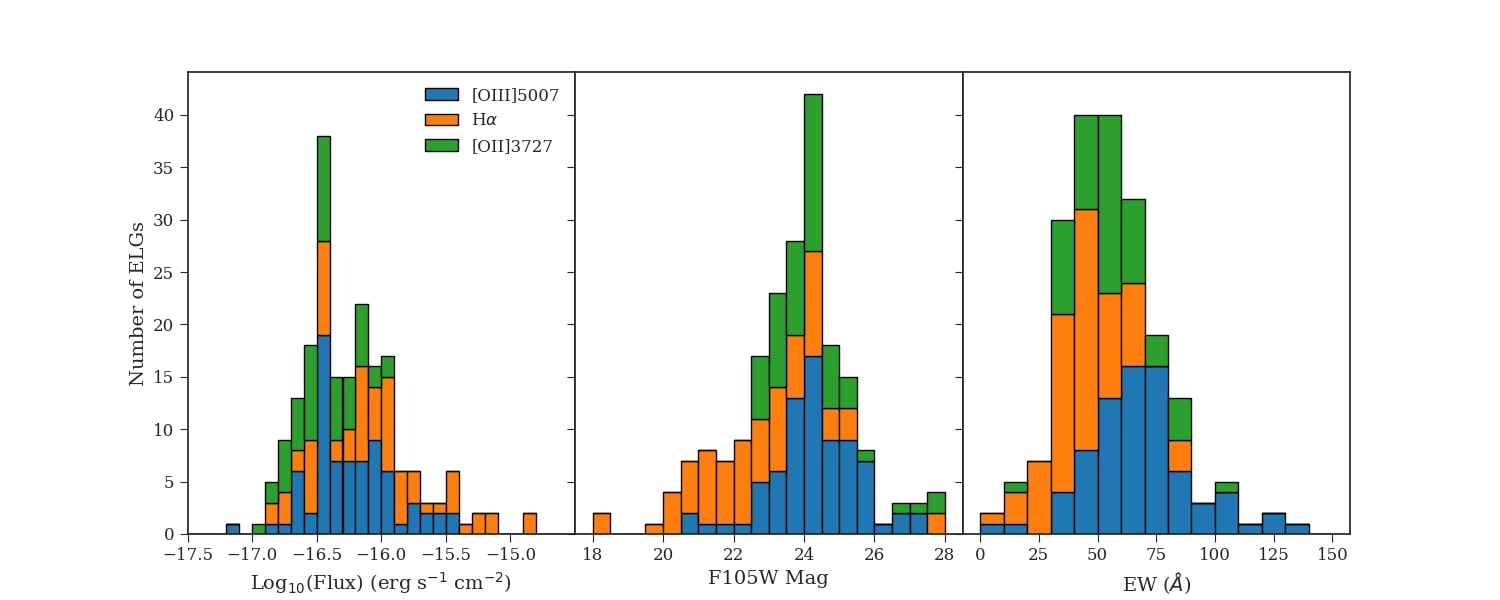}
    \caption{The distributions of ELG properties broken down by line ID. The histogram bars are colored according to the FIGS strong line ID, and bins that contain more than one type of line in FIGS have stacked bars of two or three colors. \textit{Left:} The distribution of emission line fluxes given without correction for dust extinction. \textit{Middle:} The distribution of identified ELGs by broadband F105W magnitude, in bins of 0.5 mag. \textit{Right:} The distribution of observed equivalent widths (EW) in bins of 10 \AA. The median values for each line in each quantity are given in Table \ref{tab:elgs}}
    \label{fig:elg_prop}
\end{figure}

\begin{figure}
    \centering
    \plottwo{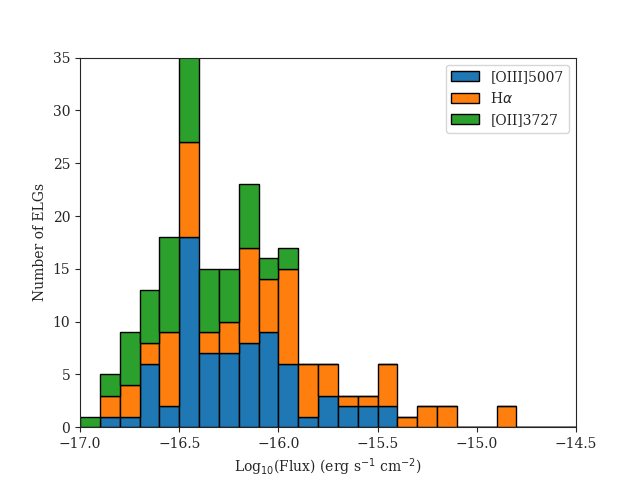}{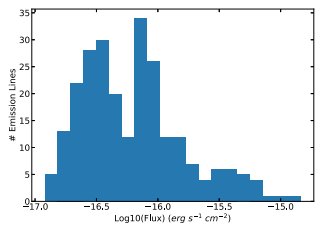}
    \caption{\textit{Left:} The ELG flux distribution from this work. The axis limits have been truncated slightly to emphasize the comparison. The full flux range can be seen in Figure \ref{fig:elg_prop} above. \textit{Right:} The ELG flux distribution from \citet{pir18}.}
    \label{fig:flux_comp}
\end{figure}

Figure \ref{fig:elg_prop} shows the distributions of some other properties of the ELG catalog. The leftmost panel gives the distribution of the observed line fluxes without correction for dust or redshift dimming for the three types of strong line emitter. This also shows the minimum line flux we were able to robustly measure, down to $10^{-17}$ erg cm$^{-2}$ s$^{-1}$. The faintest ELGs are dominated by [O\textsc{ii}]$\lambda$3727, and the brightest are dominated by the lower-redshift H$\alpha$, with [O\textsc{iii}]$\lambda \lambda$4959,5007 spanning a broad range. Figure \ref{fig:elg_prop} also shows the distribution of F105W continuum magnitudes in the middle panel, showing that we detect ELGs for F105W up to 28 mag. Finally, the rightmost panel in Figure \ref{fig:elg_prop} gives the distribution of observed equivalent widths. Given the G102 resolution of $R = 210$, detections begin to drop off significantly for EW$ < 30$ \AA. Figure \ref{fig:flux_comp} compares the 1D ELG flux distribution with the same figure from \citet{pir18}. The distributions are very similar, though \citet{pir18} find more high-flux ELGs and fewer faint-flux ELGs. This is possibly a result of the 2D method better detecting broad emission, and the 1D method detecting ELGs in fewer PAs.

\section{ELG-Overdensity Relation} \label{sec:od}

With a robust catalog of ELGs and their fluxes complete, in this section we use the catalog to probe ELG environments and explore how those environments relate to ELG properties.

\newpage

\subsection{FIGS Overdensity Catalog} \label{sec:odcat}

In \citet{pha18}, we used redshift catalogs derived from combined FIGS grism spectroscopy and broadband photometry to search for significant overdensities of galaxies in the FIGS fields. First, we divided each field into slices of redshift with $\Delta z = 0.03 \cdot (1 + z_{min})$, where $z_{min}$ is the lower bound of the redshift slice. This $\Delta z$ is based on the limiting accuracy of the \citet{pha18} photometric redshift catalogs, which enabled the most significant detections of physically associated systems. In each redshift slice, we conducted a 7th-nearest-neighbor density search for a grid of points in the field. This is defined as:

\begin{equation}
    n_7 = \frac{N}{\pi R_7^2}
\end{equation}

\noindent where $N=7$ and $R_7$ is the angular distance to the 7th-nearest galaxy in that redshift slice.

We then checked for points of significant overdensity with two different metrics. First, we calculated $\mathcal{M}$, the largest value of $n_7$ in the slice normalized to the slice's median $n_7$. We also calculated $\mathcal{S}$, the peak nearest-neighbor density in the redshift slice divided by the standard deviation of densities in the adjacent redshift slices. We counted peaks with $\mathcal{M}=10$ or $\mathcal{S}=10$ as significant detections, based on comparisons with other nearest-neighbor density searches \citep{spi12} and the values for spectroscopically identified clusters (see \citeauthor{pha18} \citeyear{pha18} for more detail). Across the 4 FIGS fields, we identified 24 such overdensities, as well as determining the $R_7$ values for individual FIGS galaxies. We make use of both the proximity to a detected overdensity and the $R_7$ distance of a galaxy to study environmental effects in the subsequent sections.

We also used the redshifts and angular separations of galaxies to determine the physical local surface density $\Sigma$ for FIGS ELGs. For the purposes of this discussion, we will use terms such that field galaxies have $\Sigma < 6$ Mpc$^{-2}$, rich fields have $6 < \Sigma < 10$, groups have $10 < \Sigma < 30$, and rich groups have $\Sigma > 30$. These definitions are adapted from those used in \citet{sob11}, where they were based on derivations from correlation length studies of galaxies in \citet{mjw96} and \citet{yan05}. We measure ELGs in the field, rich field, and group density ranges, but not in the range of rich groups or above.


\begin{figure}
\begin{tabular}{ccc}
\includegraphics[width=0.4\textwidth]{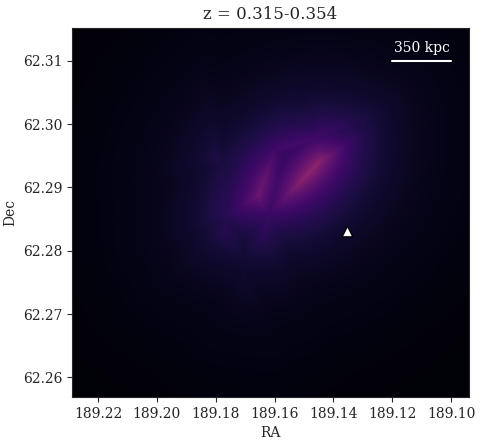} & \includegraphics[width=0.4\textwidth]{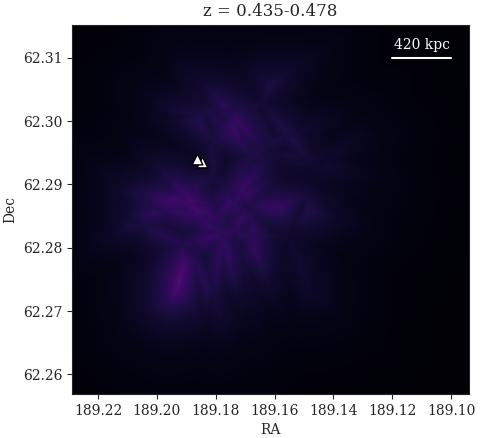} & \multirow{3}{*}{\includegraphics{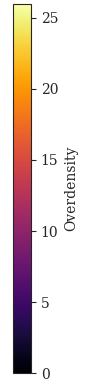}} \\  \includegraphics[width=0.4\textwidth]{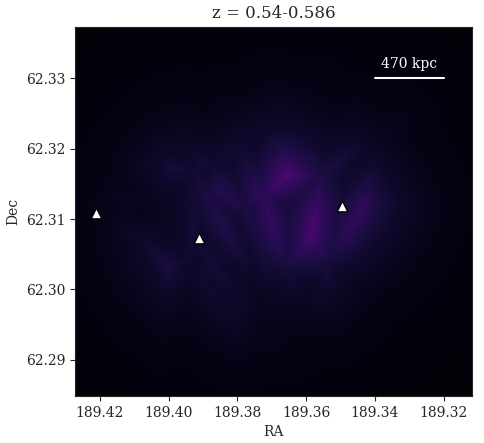} & \includegraphics[width=0.4\textwidth]{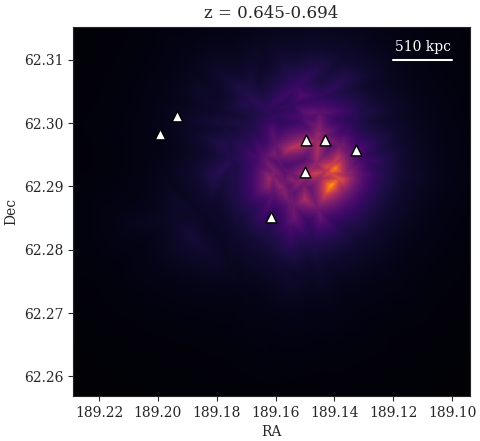} & \\ \includegraphics[width=0.4\textwidth]{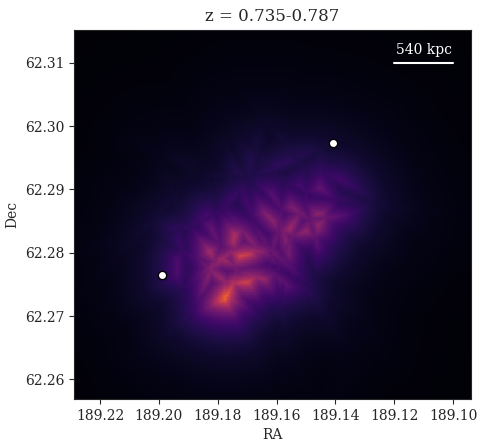} & \includegraphics[width=0.4\textwidth]{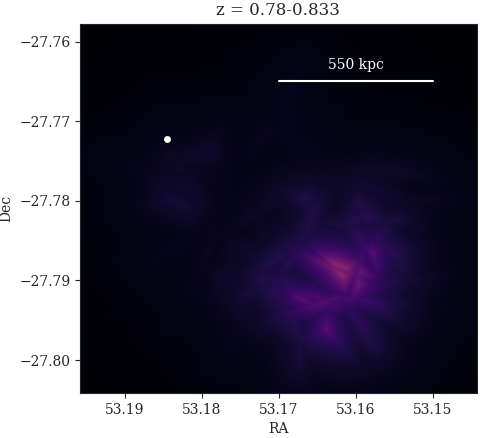} & \\ \end{tabular} 
    \caption{Example overdensities in the FIGS fields. Each image, organized from low redshift to high, shows a redshift slice where a significant overdensity was detected in \citet{pha18}, and is shaded according to the local overdensity measure, which is the ratio of the local nearest neighbor density to the median density of the redshift slice. The locations of ELGs are marked with white points, with triangles representing H$\alpha$ emitters, circles for [O\textsc{iii}] emitters, and stars for [O\textsc{ii}] emitters.}
    \label{fig:od1}

\end{figure}

\begin{figure}
\begin{tabular}{ccc}
\includegraphics[width=0.4\textwidth]{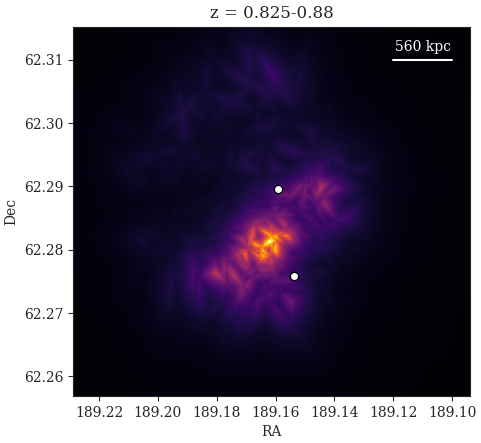} & \includegraphics[width=0.4\textwidth]{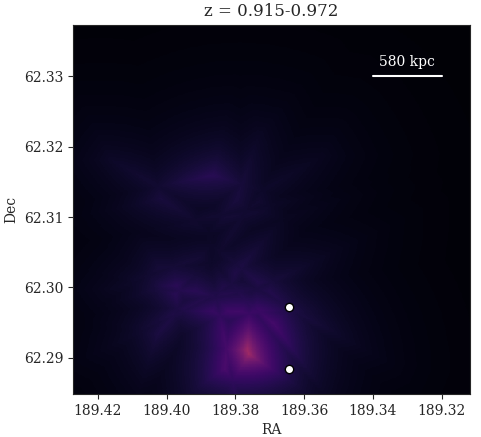} & \multirow{3}{*}{\includegraphics{colorbar.png}} \\
\includegraphics[width=0.4\textwidth]{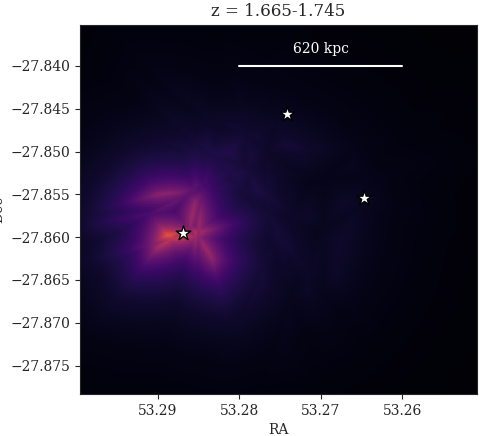} & \includegraphics[width=0.4\textwidth]{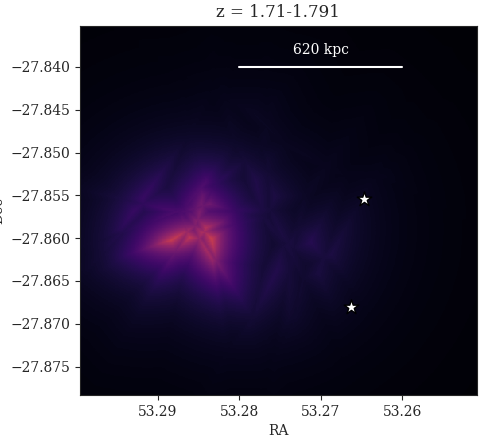} & \\ \includegraphics[width=0.4\textwidth]{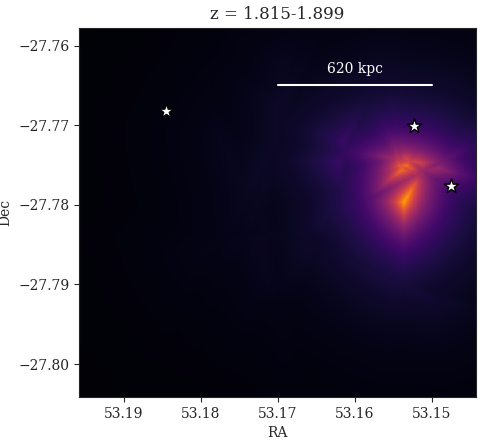} & \includegraphics[width=0.4\textwidth]{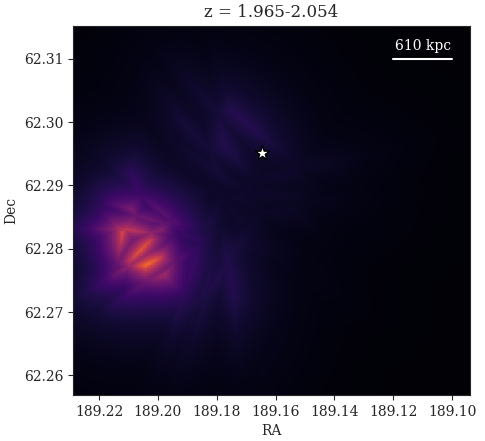} & \\
\end{tabular}
    \caption{More overdensities in the FIGS fields, including the known clusters. Each image, organized from low redshift to high, shows a redshift slice where a significant overdensity was detected in \citet{pha18}, and is shaded according to the local overdensity measure, which is the ratio of the local nearest neighbor density to the median density of the redshift slice. The locations of ELGs are marked with white points, with triangles representing H$\alpha$ emitters, circles for [O\textsc{iii}] emitters, and stars for [O\textsc{ii}] emitters.}
    \label{fig:od2}

\end{figure}

Figures \ref{fig:od1} and \ref{fig:od2} show the nearest-neighbor density plots of significant overdensities in the FIGS fields. The figures also show the locations of identified ELGs in each redshift slice, with triangles representing H$\alpha$ emitters, circles for [O\textsc{iii}] emitters, and stars for [O\textsc{ii}] emitters. There appear to be several overdensities with associated ELGs, but this does not appear to be a consistent trend visually. The spectroscopically identified clusters at $z=0.85$ and $z=1.84$ have 2 and 3 ELGs in the same redshift slice, respectively. All but one of these ELGs appear near the overdensities, but not especially near the density peaks. Neither cluster appears to show an excess of identified ELGs compared to other slices. For the $z=1.84$ cluster, this may seem in contradiction to the results of \citet{tra10}, which found increased star formation activity near a cluster core at similar redshift. It's possible this is because the relatively faint [O\textsc{ii}] is simply less complete compared to the lower-redshift sources, or because we do not find many galaxies with surface densities of $\Sigma \geq 20$ Mpc$^{-2}$, the density at which \citet{tra10} begin to find the increased fraction of star-forming galaxies. We discuss the SFRs themselves and their implications in \S5.3-5.5.

\subsection{The $R_7$ Distribution}

In order to systematically study a possible relationship between strong line emission and galaxy environment, we first looked at the $R_7$ distance for both ELGs and regular galaxies. If ELGs have a preferred relationship with overdensities, then the distribution of $R_7$ distances could be distinct from non-emitting galaxies, since, for example, a preference for ELGs to be close to overdensities should result in a distribution that peaks more at low $R_7$.

\begin{figure}
\plotone{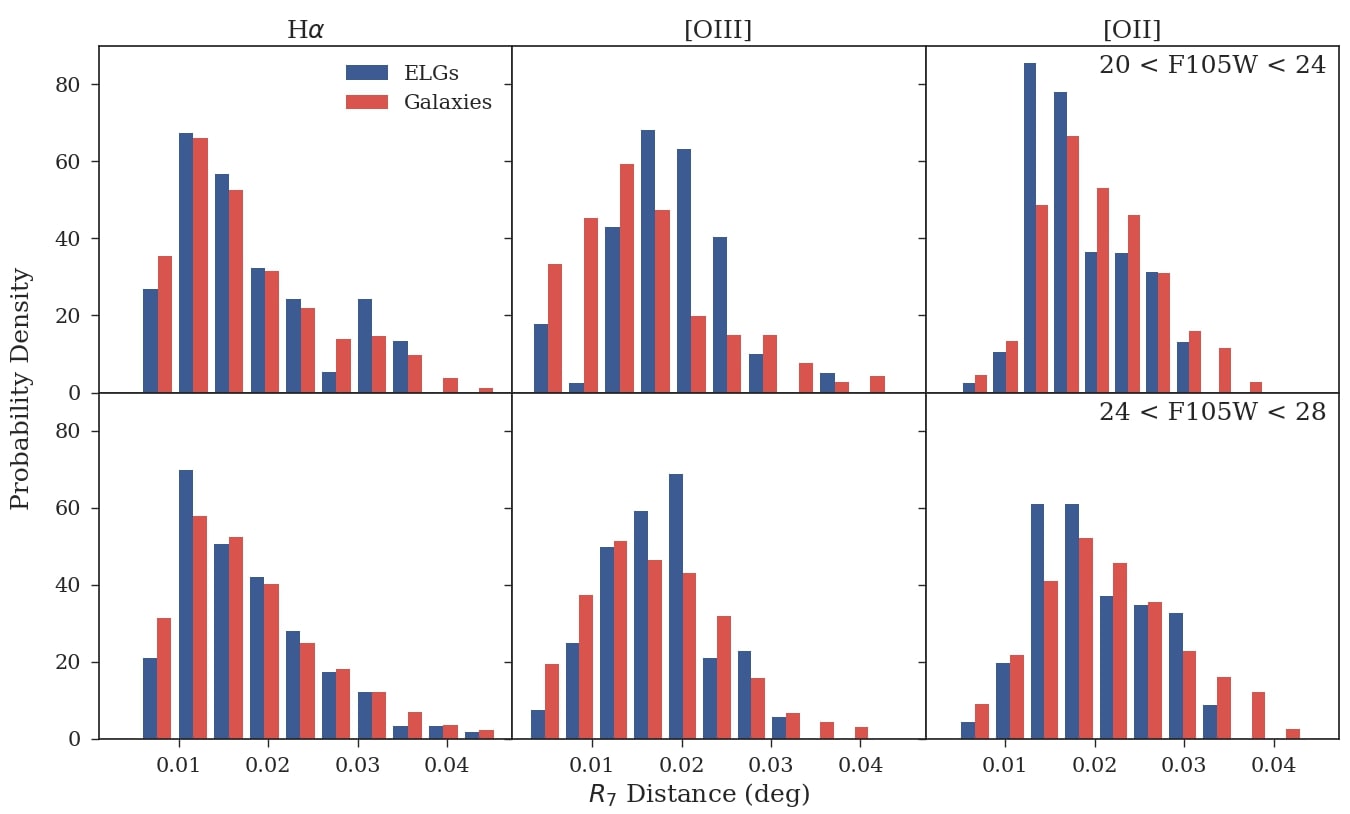}
\caption{The probability density distributions of $R_7$ distances, separated into bins of redshift and F105W continuum magnitude, for ELGs (blue) and all galaxies (red) in a given magnitude-redshift bin. The first column uses $0.3 < z < 0.8$, corresponding to H$\alpha$ emission. The middle column uses $0.8 < z < 1.3$ ([O\textsc{iii}]$\lambda\lambda 4959,5007$), and the right column uses $1.3 < z < 2.1$ ([O\textsc{ii}]$\lambda3727$). The top row compares ELGs and galaxies with $ 20 < F105W < 24$ mag (bright). The bottom row compares ELGs and galaxies with $ 24 < F105W < 28$ mag (faint). For each distribution pair, we applied a two-sample K-S test to determine whether the distributions differed significantly. Both the bright and the faint H$\alpha$ distributions are indistinguishable from the distributions of galaxies. The distributions of the [O\textsc{iii}]$\lambda\lambda 4959,5007$ emitters do differ significantly ($p=3\cdot 10^{-7}$ for the bright distribution, $p=0.02$ for the faint), with the [O\textsc{iii}]$\lambda\lambda 4959,5007$ emitters found preferentially at middling $R_7$ values as opposed to low $R_7$. The [O\textsc{ii}]$\lambda3727$ distribution differs significantly only in the bright sample (p$=0.02$).}
\label{fig:r7}
\end{figure}

Figure \ref{fig:r7} shows the probability density distributions of $R_7$ distances for ELGs compared to the whole set of galaxies in our redshift catalog. The distributions are broken down into six subsamples, in order to make meaningful comparisons of distance and stellar mass: first, by redshift ranges corresponding to the three strong emitters, and then by bright and faint F105W continuum magnitudes as a proxy for mass. To judge the significance of the differences between a given pair of distributions, we applied a two-sample Kolmogorov-Smirnov (KS) test, a statistical test to determine whether two underlying one-dimensional probability distributions differ, to each subsample pair. The test produces a p-value determined by the sizes and differences of the two distributions, and this p-value gives the level of significance at which the two may be considered distinct. A lower p-value corresponds to a more significant determination that the two distributions are different. 

This test showed no significant difference in the $R_7$ distribution of H$\alpha$ emitters compared to other galaxies in either the bright or faint bins. For [O\textsc{iii}]$\lambda \lambda$4959,5007 emitters, the test does find a significant difference in distributions for both the bright ($p = 3\cdot10^{-7}$) and faint ($p = 0.02$) bins, with [O\textsc{iii}]$\lambda \lambda$4959,5007 emitters having a higher probability of appearing at mid-range $R_7$ distances compared to other galaxies. For  [O\textsc{ii}]$\lambda$3727 emitters, the test finds a significant difference only in the bright bin ($p = 0.002$). This measurement for [O\textsc{iii}] supports previous studies that find line-emitters preferentially at intermediate distances around clusters \citep{dar14} at $z \simeq 1$. We explore this result in more detail in \S5.5.


\newpage
\subsection{Measuring Star Formation}

Studying the $R_7$ distribution by itself gives insight into only the relationship between the locations of ELGs and of overdensities, while ignoring the other properties of the ELGs. With the flux catalog, we were also able to investigate how an ELG's environment might influence its emission line luminosity and recent star formation rate.



To account for the effects of dust extinction in measuring the SFR, we used a dust calibration developed by \citet{sob12} using rest-frame \textit{u}-\textit{z} colors. The calibration was developed and tested using H$\alpha$ and [O\textsc{ii}] emitters at $z=0.1$ and $z=1.47$. It is given by

\begin{equation}
    A_{\text{H}\alpha} = -0.092(u-z)^3 + 0.671(u-z)^2 - 0.952(u-z) + 0.875
\end{equation}

\noindent \citet{sob12} find that this relation holds across redshift epochs, covering most of the redshift range of our sample and for both kinds of emitters. To convert the $A_{\text{H}\alpha}$ calculation to $A_{[OIII]}$ and $A_{[OII]}$, we applied the \citet{cal00} reddening law. For the few objects for which one of rest-frame \textit{u} or \textit{z} was unavailable, we assigned the median reddening value from the rest of the sample. We also measured $A_{H\alpha}$ using the stellar mass dust parameterization developed in \citet{gb10} (see \S5.4 for discussion of stellar masses). This typically produced similar dust measures to the color calibration, with a divergence at masses greater than $10.5 \log(M_{\odot})$, as was also observed in \citet{ram17}. Such objects make up a very small fraction of our ELG sample, so this doesn't change any overall trends. 

We calculated SFRs for the ELGs using the following equations:

\begin{equation}
    \text{SFR}_{\text{H}\alpha}\text{ ($M_{\odot}$ yr$^{-1}$)} = 7.9 \times 10^{-42} L(\text{H}\alpha)\text{ (erg s$^{-1}$)} \label{eq:ha}
\end{equation}

\begin{equation}
    \text{SFR}_{[\text{O\textsc{ii}}]}\text{ ($M_{\odot}$ yr$^{-1}$)} = 1.4 \pm 0.4 \times 10^{-41} L(\text{[O\textsc{ii}]})\text{ (erg s$^{-1}$)} \label{eq:o2}
\end{equation}

\begin{equation}
    \text{SFR}_{[\text{O\textsc{iii}}]}\text{ ($M_{\odot}$ yr$^{-1}$)} = 6.4 \pm 4.0 \times 10^{-42} L(\text{[O\textsc{iii}]})\text{ (erg s$^{-1}$)} \label{eq:o3}
\end{equation}

\noindent Equations \ref{eq:ha} and \ref{eq:o2} are calibrations from \citet{ken98}, derived with a Salpeter IMF. Equation \ref{eq:o3} was derived by \citet{str09} using [O\textsc{iii}]-H$\alpha$ ratios from star-forming galaxy knots where both emission lines were present.


\begin{figure}
\plotone{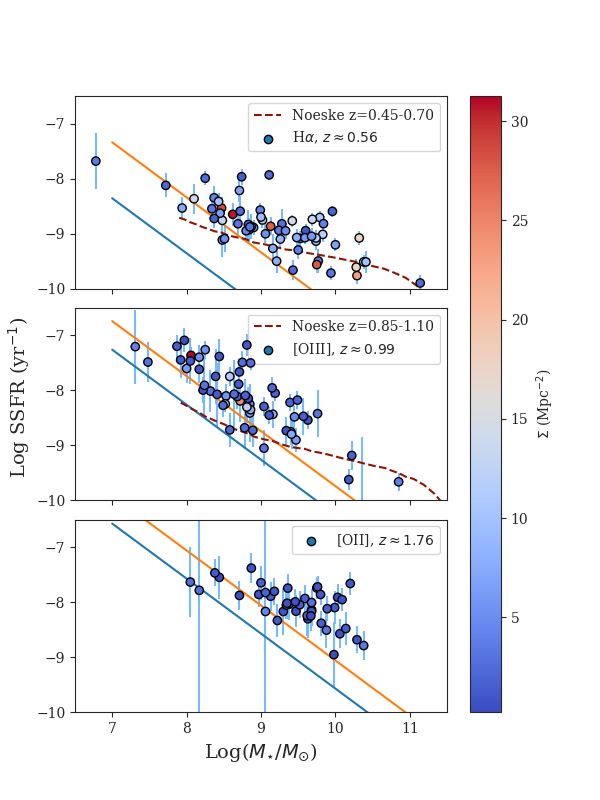}
\caption{The specific star formation rate (SSFR) as a function of the stellar mass. The ELGs are given by colored circles (H$\alpha$, top panel; [O \textsc{iii}]$\lambda\lambda 4959, 5007$, middle; [O\textsc{ii}]$\lambda 3727$, bottom), and are shaded by their value of $\Sigma$, the local density of galaxies (see \S5.5 for description). The median $z$ for each panel's FIGS subsample is given in the panel legend. Each panel also shows two lines of completeness (blue and orange solid lines), derived from the limiting line flux we measured (see Figure \ref{fig:elg_prop}), and the minimum and maximum possible redshifts for each line. There is no clear trend between $\Sigma$ and a galaxy's position on the SSFR-mass relation. The red dashed curves are the best-fit staged-tau models from \citet{noe07}. Our ELGs typically sit at higher SSFR for a given stellar mass compared to the models at comparable redshift, but this is likely due to the flux limitations of our sample.} \label{fig:ssfr_mass}
\end{figure}

\subsection{Environment and the SSFR-Mass Relation}

We obtained stellar masses for our ELG sample by applying our EAZY SED catalogs to the SED fitting code FAST \citep{kri09}, using a \citet{cha03} initial mass function, \citet{bc03} templates, and an exponential star formation history. We used spec-zs for the fits where high-quality matching redshifts were available in our compilation, and used the best-fit photometric redshift otherwise. We checked the results for the GS1 field against the GOODS-South catalogs compiled by \citet{san15}, which largely exclude our GS2 parallel field. For the galaxies with existing measurements, we found our mass results consistent with the \citet{san15} catalogs, with a median difference of less than 0.1 dex. For the other fields, we checked against matching masses from \citet{sk14}, and found a similar level of agreement.

With the stellar masses calculated, we were able to determine each ELG's SFR per stellar mass, or specific star formation rate (SSFR). The relation between SSFR and mass for star forming galaxies is typically called the galaxy main sequence, and it suggests an evolution of star formation with redshift and stellar mass \citep{noe07}. In star forming galaxies, as redshift decreases, ongoing star formation builds up increased stellar mass, and as this happens SSFRs decline as the galaxies exhaust their supplies of gas.

This smooth relation doesn't account for cases of rapid quenching, and does not address the influence of environment on how galaxies evolve. In Figure \ref{fig:ssfr_mass}, we show the SSFR as a function of the stellar mass, and compare our ELGs to the results from \citet{noe07}, who measured this in four redshift bins up to $z = 1.1$. We show the best-fit staged-tau models of star formation history from \citet{noe07} in the redshift bins most closely matching our H$\alpha$ and [O\textsc{iii}] emitters (the [O\textsc{ii}] sample is at too high a redshift), and find that our ELGs typically have higher SSFR for a given stellar mass. 

This may be due to the luminosity limit of our sample. Since the limiting luminosity is redshift dependent, we determined the lower-bound SSFR based on the lower and upper redshift limits of each ELG sample. We show these lines in Figure \ref{fig:ssfr_mass}. For [O\textsc{iii} and [O\textsc{ii}], the limiting SSFRs are relatively high, indicating that we are likely sampling the upper region of the galaxy main sequence.


We were still able to investigate the possible environmental effects on the main sequence, especially for H$\alpha$, where we probe the main sequence most closely. In Figure \ref{fig:ssfr_mass}, we also show each galaxy's local surface density $\Sigma$. There is no significant relationship between either $\Sigma$ and the stellar mass or $\Sigma$ and position on the SSFR-mass relation to the densities probed, which at most get only as dense as galaxy groups ($10 < \Sigma < 30$, as defined in \S5.1.). The same holds for the ELGs' $R_7$ measurements. This suggests that environmental effects do not play a systematic role in either quenching or triggering star formation, since they do not appear to disrupt the smooth star formation relation of the main sequence. 

\subsection{Line Luminosity and Clustering}

\begin{figure}
\plotone{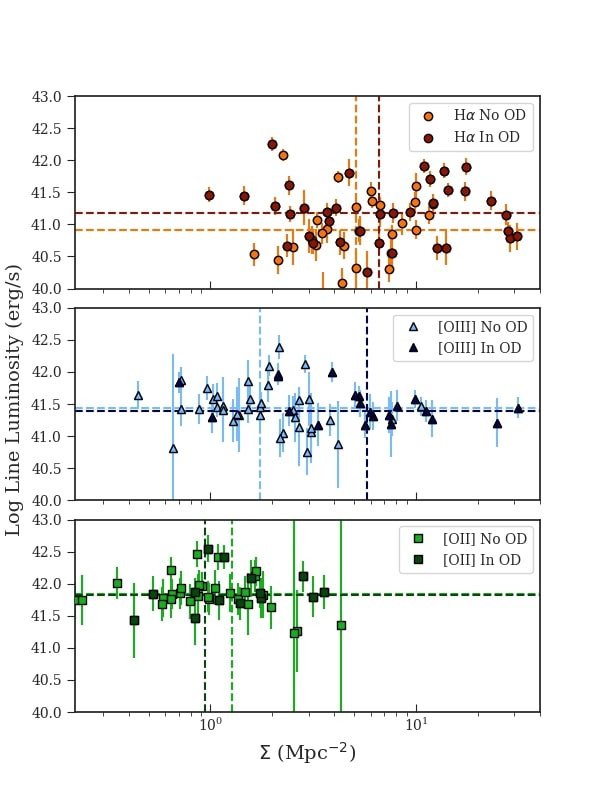}
\caption{The line luminosity as a function of surface density for FIGS ELGs. The ELGs are given by orange circles (H$\alpha$, top panel), blue triangles ([O \textsc{iii}]$\lambda\lambda 4959, 5007$, middle), and green Xs ([O\textsc{ii}]$\lambda 3727$, bottom). Each panel contains two subsamples: emitters found in redshift slices without a significant overdensity detection ("No OD"; lighter colors), and emitters found in slices with a significant detection ("In OD"; darker colors). The median luminosity and $\Sigma$ values for each subsample are given by the horizontal and vertical dashed lines. This shows a substantial difference in the locations of [O\textsc{iii}] emitters depending on the proximity of an overdensity: in redshift slices with an overdensity, the [O\textsc{iii}] emitters are much more likely to be found at densities corresponding to group outskirts.}
\label{fig:l_sig2}
\end{figure}

We also studied the relationship between emission line strength and galaxy clustering more directly. First, we used the galaxies' redshifts and angular separations to compute the local physical surface density $\Sigma$ in units of Mpc$^{-2}$ for each ELG. Then, we split the sample of ELGs into two subsamples: those located in a redshift slice where a significant overdensity is detected ("In OD"), and those in a redshift slice with no overdensity detection ("No OD"). By looking at these subsamples, we could check if ELG properties differ for galaxies near a range of peak densities. This also gives us a subsample to compare directly to studies focused only on known clusters. This result can be seen in Figure \ref{fig:l_sig2}, which shows the line luminosity as a function of $\Sigma$ for H$\alpha$, [O\textsc{iii}], and [O\textsc{ii}] emitters. In each panel, the horizontal dashed lines give the median line luminosity for each subsample, and the vertical dashed lines give the median $\Sigma$. We inspected the SFR-$\Sigma$ and SSFR-$\Sigma$ relations as well (see Figure \ref{fig:sfr_sig}), and the distributions remained essentially unchanged for each emitter. Thus, for this discussion we will refer simply to the L-$\Sigma$ relation shown in Figure \ref{fig:l_sig2}, as that requires the fewest additional assumptions.


\begin{figure}
    \centering
    \plottwo{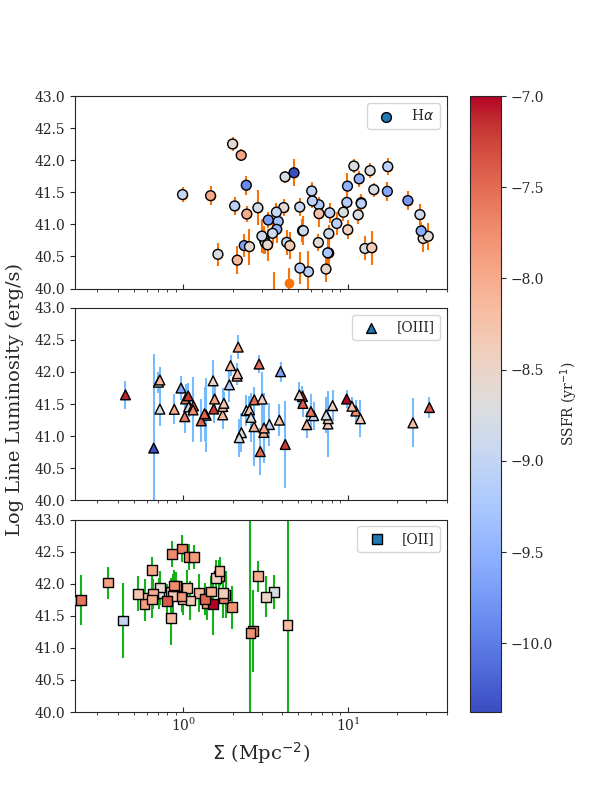}{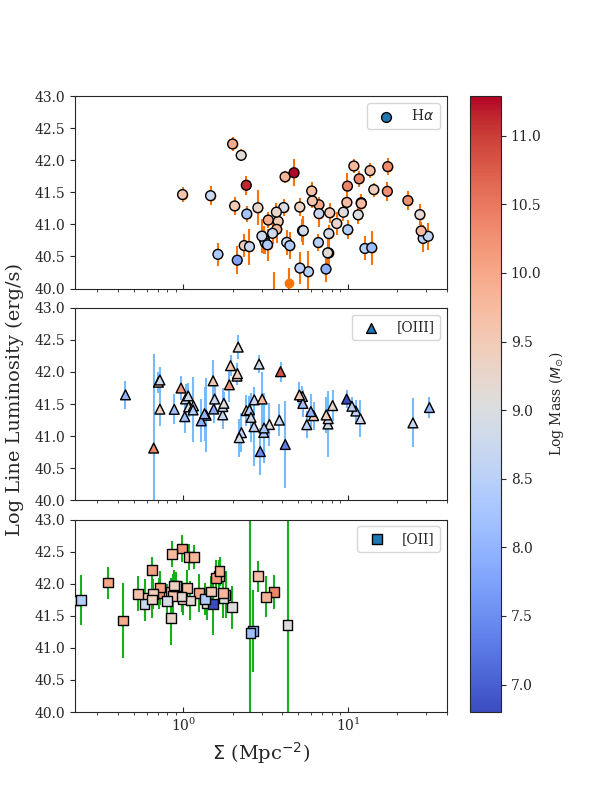}
    \caption{Left: The L-$\Sigma$ distribution colored by SSFR. Right: The L-$\Sigma$ distribution colored by stellar mass.}
    \label{fig:other_comp}
\end{figure}

\begin{figure}
\plotone{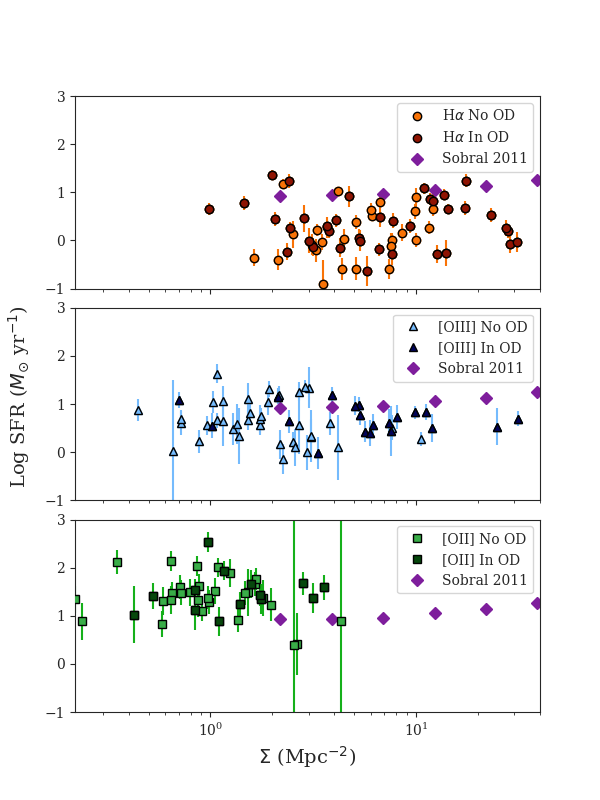}
\caption{The SFR-$\Sigma$ relation. The ELGs are given by orange circles (H$\alpha$, top panel), blue triangles ([O \textsc{iii}]$\lambda\lambda 4959, 5007$, middle), and green Xs ([O\textsc{ii}]$\lambda 3727$, bottom). Each panel contains two subsamples: emitters found in redshift slices without a significant overdensity detection ("No OD"; lighter colors), and emitters found in slices with a significant detection ("In OD"; darker colors). Median SFR bins from \citet{sob11} are shown with purple diamonds.}
\label{fig:sfr_sig}
\end{figure}

The H$\alpha$ L-$\Sigma$ distribution (top panel) shows little indication of a preferred relationship with density. Even in the OD slices, H$\alpha$ emitters are found at a range of local densities, and with a range of luminosity values. The median luminosity for emitters near overdensities is 0.3 dex higher than for those in non-OD slices, which is about twice the typical error size for the H$\alpha$ emitters. Figure \ref{fig:other_comp} shows the same L-$\Sigma$ relation with the points shaded by SSFR and stellar mass, but no clear trends with density emerge.

In Figure \ref{fig:sfr_sig}, we compared the derived H$\alpha$ SFRs from this luminosity sample to a narrowband-selected sample of star-forming (SFR $ > 3 M_{\odot}$ yr$^{-1}$) H$\alpha$ emitters at $z = 0.845$ \citep{sob11}, using the same SFR diagnostic. Our distribution differs from the result in \citet{sob11}, which shows SFR increasing with density up to $\Sigma \simeq 50$ Mpc$^{-2}$. The discrepancy may be at least partially explained by their selection of SFR $ > 3 M_{\odot}$ yr$^{-1}$ emitters, which would exclude much of our sample. SFR measurements of FIGS ELGs go down to $0.1 M_{\odot}$ yr$^{-1}$, but our sample probes to lower stellar mass, yielding comparable SSFRs. Nevertheless, this does limit the utility of a comparison with the \citet{sob11} results. \citet{dar14} conducted a similar study down to a limit of SFR $ > 1.5 M_{\odot}$ yr$^{-1}$, and finds only a small difference in median SFR between field and cluster galaxies. However, \citet{dar14} also find that at intermediate densities, comparable to the rich field or group densities described earlier, a higher fraction of galaxies exhibit star formation compared to fields and rich clusters.

To get a broader sense of environmental effects at different redshifts, we need to look at non-narrowband studies. \citet{gru11} found a weak correlation between overdensity and \textit{U-B} color, used as a proxy for star formation, using data from the POWIR and DEEP2 surveys. Their study showed declining star formation with increasing density at $0.4 \leq z \leq 0.7$, roughly matching the redshift range of our H$\alpha$ emitters, which becomes a flat relation for $0.85 \leq z \leq 1$. In their paper, they do not attempt to convert the color to SFR or SSFR, but in \citet{gru11c} the change in observed color is shown to correlate with a half-dex decrease in SSFR at $1.5 < z < 2$. \citet{gru11} also measure these effects in terms of overdensity, not physical density, so a direct comparison with our results and that of \citet{sob11} is not possible. However, their largest overdensity bin corresponds to the value of our overdensity cutoff; in this case, one might expect to see a decline in H$\alpha$ line luminosity among those galaxies marked "In OD," but no such decline is apparent. The result of \citet{gru11} was limited to stellar masses down to only $\log (M_{\star}/M_{\odot}) > 10.25$, so they may not have detected the low-luminosity scatter we observe at low $\Sigma$, while observing a decline in high SFR as $\Sigma$ increases. 

\citet{sco13} studied NUV-continuum-derived SFRs versus density in redshift slices up to $z < 3$ in COSMOS, finding a flat SFR-$\Sigma$ relationship for $0.8 < z < 2$. \citet{sco13} studied the SFR-$\Sigma$ relation in terms of density percentiles, of which the highest encompassed $\Sigma > 10$ Mpc$^{-2}$ and the lowest $\Sigma < 0.1$ Mpc$^{-2}$. At $z < 1$, they measure a flat relationship up to $\Sigma$ of a few for the lower and medium density percentiles, after which the SFR declines with increasing density in the highest percentiles. At $0.35 < z < 0.6$, the redshift bin most comparable to our H$\alpha$ sample, the decline is notable in only the highest-density bin, where the median SFR drops from $1.5 M_{\odot}$ to $0.3 M_{\odot}$. We do not observe this drop in the line luminosity (or in dust-corrected SFRs), but the high-density percentile from \citet{sco13} includes densities up to $\sim100$ Mpc$^{-2}$, and perhaps the drop in SFR is concentrated in these very high density sources. If so, it could be the case that we do not detect ELGs at such high density because their star formation has dropped too low, leading to fainter (or no) line emission. The trends from \citet{sco13}, combined with the results of \citet{dar14}, could suggest a transition period from peak star formation and merger interactions at $z \simeq 2$ \citep{md14} to the local universe. After the merger peak, higher-density environments may have already quenched star formation in local galaxies through strangulation or ram-pressure stripping \citep{muz12, muz14}, depleting their reserves of star-forming material and increasingly relegating star formation to intermediate and field densities. This could be the case with our H$\alpha$ sample, if the lack of high-$\Sigma$ emitters is due to  the depletion of star formation at high density.

The [O\textsc{iii}]$\lambda\lambda 4959,5007$ L-$\Sigma$ distribution (middle panel) shows an essentially flat relationship, with high scatter at the lowest densities. In slices with an overdensity, the emitters we find are much more likely to occupy densities in the range $5 < \Sigma < 15$, with 12 out of 20 emitters in OD slices found in this region. The median $\Sigma$ is a factor of a few higher for [O\textsc{iii}] emitters in overdensity slices compared to those not near overdensities. This is distinct from the H$\alpha$ emitters, which don't seem to have a distinct density preference when near overdensities. 

The density where the [O\textsc{iii}] emitters are preferentially located corresponds to the rich fields and galaxy groups at the outskirts of a denser cluster, corroborating what we see in Figure \ref{fig:r7}. This result also matches the findings in \citet{gru11} and \citet{sco13}, as we find a flat SFR-density relationship at $0.8 < z < 1.3$. We do not see the higher SFR at intermediate density that \citet{dar14} measured, but our results do corroborate their finding of a higher fraction of star-forming galaxies at those densities. Compared to the H$\alpha$ distribution, which shows emitters at all density ranges near overdensities, this could suggest an evolution with redshift in the preferred locations of group star forming galaxies near overdensities, which are found nearer to rich group densities among the FIGS H$\alpha$ emitters.


The [O\textsc{ii}]$\lambda3727$ SFR-$\Sigma$ distribution (bottom panel) shows no relationship, with all the emitters found at low $\Sigma$. This is likely due in part to the limits of our overdensity search near $z \simeq 2$, where our sample of fainter galaxies is less complete (see \S5.4 for further discussion). One can see this effect in the range of SFRs calculated for the [O\textsc{ii}] emitters, which is restricted to much higher star formation than the other two samples. Of the 24 overdensity candidates, only 4 are in the redshift range where we might find [O\textsc{ii}], and these are of lesser significance and based on fewer galaxies compared to the overdensity candidates at lower redshift. These caveats aside, this could suggest that at higher redshift, field galaxies exhibit higher star formation, at least among the brightest galaxies. \citet{pat09} found in a study of $z \simeq 0.8$ galaxies with $\log(M_{\star}/M_{\odot}) > 10$ that specific star formation (SFR per stellar mass) declined with increasing density. Since the limits of our completeness at this redshift selects more massive galaxies, this could indeed explain our findings.

\subsection{Line Emission and Galaxy Pairs}

We also investigated the behavior of ELGs with close companion galaxies, in order to study overdensities and environmental effects on a smaller scale. If nebular emission and related star formation are triggered by interactions with companion galaxies \citep{ken87, alo04}, then we could observe a difference in the number of nearby galaxies between ELGs and galaxies. \citet{ell10} studied the effects on environment of interacting galaxy pairs selected from SDSS DR4, finding a small increase in SSFR for the closest pairs at low $\Sigma$ relative to both more distant pairs and pairs found at higher galaxy densities. Their distance criterion for identifying a pair required a projected distance of $R_p < 80 \times h^{-1}_{70}$ kpc between the two galaxies. Using this projected distance as the range for possible companions, we find that the fractions of ELGs (31\%) and non-ELGs (32\%) that form a near pair are essentially the same. \citet{koc12} used a much narrower allowable pair range (12 kpc) to search for interactions near AGN hosts. Applying this much stricter cut yields a 3\% pair rate in both ELGs and non-ELGs, suggesting that line emission and star formation are not necessarily directly connected to the presence of a nearby companion. 







\section{Conclusions}

In this paper, we used deep NIR slitless spectroscopy to conduct an automated search for emission line galaxies. Using our continuum-subtracted peak-finding technique, we detected and identified 208 H$\alpha6563$, [O\textsc{iii}]$\lambda\lambda$4959,5007, and [O\textsc{ii}]$\lambda$3727 emitters in the four FIGS fields. For these emitters, we provide a robust catalog with integrated line fluxes, flux errors, line-derived redshifts, and observed equivalent widths. We measure line fluxes down to $10^{-17}$ erg cm$^{-2}$ s$^{-1}$ for objects with continuum magnitudes up to F105W $< 28$ mag. We compare line-derived redshifts to high-quality spectroscopic redshifts and find that 80\% of ELG redshifts match the spec-zs within $\Delta z = \pm 0.0025$, the width of the WFC3-G102 grism element. We find no dependence of this redshift accuracy on galaxy size.

We use the flux catalog to derive SFRs and the local surface densities of galaxies, which we use to search for trends in the SFR-density relation. We find that [O \textsc{iii}] emitters are preferentially found at intermediate densities in the outer regions of galaxy groups, as shown in Figure \ref{fig:r7}, corroborating a finding at similar redshifts. When placing our sample on the SSFR-mass relation, we find higher SSFR per stellar mass compared to other studies at comparable redshift, though this is largely explained by limits on measured line flux. We find that SFR has no significant dependence on increasing local galaxy surface density for $0.3 < z < 0.8$ H$\alpha$ emitters and for $0.8<z<1.3$ [O\textsc{iii}] emitters, as shown in Figure \ref{fig:l_sig2}. We find no indication that environment influences a galaxy's location in this relation. A study of close galaxy pairs finds that ELGs are not more or less likely to have a close companion than non-ELGs. We compare our results with other environment studies after the peak in cosmic star formation at $z \simeq 2$ \citep{md14}, which show a variety of possible relations across different redshifts. We observe a difference in the preferred location of rich field and group ELGs near overdensities, from a preference for rich field densities at $z \simeq 1$ to no preference between field and group densities at $z \simeq 0.5$. We do not find ELGs at the high surface densities common to righ groups or clusters, which could be due to low star formation at those densities. \\

We would like to thank the referee for many thoughtful comments and helpful suggestions. This work is based on observations made with the NASA/ESA Hubble Space Telescope, obtained from the Data Archive at the Space Telescope Science Institute, which is operated by the Association of Universities for Research in Astronomy, Inc., under NASA contract NAS 5-26555. These observations are associated with program \#13779. Support for program \#13779 was provided by NASA through a grant from the Space Telescope Science Institute, which is operated by the Association of Universities for Research in Astronomy, Inc., under NASA contract NAS5-26555.  A.C. acknowledges the grants ASI n.I/023/12/0 ``Attivit relative alla fase B2/C per la missione Euclid'' and PRIN MIUR 2015 ``Cosmology and Fundamental Physics:illuminating the Dark Universe with Euclid.''

\appendix

\section*{Appendix A: Complete Line Flux Catalog}

\startlongtable
\begin{deluxetable*}{ccccccccc}
\label{tab:flux}
\tablecolumns{8}
\tablehead{
\colhead{} \vspace{-0.2cm} &
\colhead{} &
\colhead{RA} &
\colhead{Dec} &
\colhead{F105W} &
\colhead{} &
\colhead{Flux ($10^{-18}$}  &
\colhead{} &
\colhead{EW}  \\
\colhead{Field} \vspace{-0.6cm} &
\colhead{ID} &
\colhead{(Deg)} &
\colhead{(Deg)} &
\colhead{(AB Mag)} &
\colhead{Line} &
\colhead{ erg s$^{-1}$ cm$^{-2}$)} &
\colhead{$z_{line}$} &
\colhead{(\AA)} \\
}
\startdata
GN1 & 1134 & 189.167313 & 62.306263 & 24.4 & H$\alpha$ & $17.1\pm3.7$ & 0.636 & $41\pm8$\\
GN1 & 1144 & 189.139786 & 62.305721 & 22.6 & H$\alpha$ & $66.3\pm12.6$ & 0.557 & $37\pm6$\\
GN1 & 1225 & 189.201447 & 62.304108 & 22.4 & H$\alpha$ & $116.5\pm88.2$ & 0.636 & $55\pm40$\\
GN1 & 1289 & 189.172318 & 62.302406 & 23.2 & H$\alpha$ & $59.2\pm7.7$ & 0.529 & $41\pm5$\\
GN1 & 1297 & 189.156693 & 62.302139 & 24.2 & H$\alpha$ & $42.5\pm11.6$ & 0.384 & $57\pm15$\\
GN1 & 1339 & 189.193359 & 62.30109 & 22.9 & H$\alpha$ & $92.7\pm21.5$ & 0.672 & $55\pm12$\\
GN1 & 1344 & 189.182663 & 62.301079 & 23.8 & [O\textsc{iii}] $\lambda\lambda4959,5007$ & $52.3\pm11.2$ & 1.014 & $69\pm13$\\
GN1 & 1354 & 189.178833 & 62.300762 & 24.7 & [O\textsc{iii}] $\lambda\lambda4959,5007$ & $22.6\pm10.1$ & 1.09 & $52\pm22$\\
GN1 & 1413 & 189.134064 & 62.299328 & 22.7 & [O\textsc{iii}] $\lambda\lambda4959,5007$ & $121.3\pm13.3$ & 1.013 & $55\pm6$\\
GN1 & 1458 & 189.199326 & 62.29826 & 23.5 & H$\alpha$ & $32.5\pm7.9$ & 0.647 & $48\pm12$\\
GN1 & 1485 & 189.143372 & 62.297356 & 21.6 & H$\alpha$ & $91.3\pm15.6$ & 0.684 & $27\pm4$\\
GN1 & 1494 & 189.149567 & 62.297413 & 24.2 & H$\alpha$ & $28.3\pm6.2$ & 0.678 & $44\pm9$\\
GN1 & 1497 & 189.15683 & 62.296238 & 20.6 & H$\alpha$ & $369.6\pm75.1$ & 0.554 & $42\pm8$\\
GN1 & 1499 & 189.140961 & 62.297306 & 24.6 & [O\textsc{iii}] $\lambda\lambda4959,5007$ & $47.4\pm8.8$ & 0.799 & $62\pm11$\\
GN1 & 1508 & 189.150726 & 62.297047 & 22.9 & H$\alpha$ & $75.1\pm11.1$ & 0.711 & $42\pm6$\\
GN1 & 1539 & 189.132751 & 62.295826 & 21.1 & H$\alpha$ & $122.8\pm29.8$ & 0.684 & $27\pm6$\\
GN1 & 1583 & 189.164795 & 62.295155 & 23.0 & [O\textsc{ii}] $\lambda3727$ & $113.0\pm19.5$ & 2.049 & $86\pm15$\\
GN1 & 1589 & 189.153412 & 62.295105 & 25.4 & [O\textsc{iii}] $\lambda\lambda4959,5007$ & $39.8\pm8.6$ & 0.964 & $59\pm13$\\
GN1 & 1610 & 189.15947 & 62.294628 & 23.2 & H$\alpha$ & $36.3\pm10.4$ & 0.604 & $40\pm11$\\
GN1 & 1640 & 189.186539 & 62.293983 & 23.6 & H$\alpha$ & $55.0\pm9.0$ & 0.456 & $49\pm7$\\
GN1 & 1647 & 189.184692 & 62.29356 & 21.1 & H$\alpha$ & $441.0\pm97.9$ & 0.451 & $61\pm13$\\
GN1 & 1681 & 189.161194 & 62.293125 & 26.0 & [O\textsc{iii}] $\lambda\lambda4959,5007$ & $45.3\pm7.8$ & 1.022 & $67\pm11$\\
GN1 & 1715 & 189.149933 & 62.292282 & 24.2 & H$\alpha$ & $25.3\pm7.5$ & 0.68 & $38\pm12$\\
GN1 & 1734 & 189.139267 & 62.291878 & 23.5 & [O\textsc{iii}] $\lambda\lambda4959,5007$ & $14.6\pm5.4$ & 1.284 & $19\pm7$\\
GN1 & 1747 & 189.1521 & 62.29171 & 23.7 & [O\textsc{iii}] $\lambda\lambda4959,5007$ & $50.9\pm9.8$ & 1.218 & $59\pm11$\\
GN1 & 1750 & 189.173691 & 62.291481 & 21.8 & H$\alpha$ & $98.8\pm18.8$ & 0.486 & $34\pm6$\\
GN1 & 1756 & 189.151215 & 62.29155 & 24.2 & [O\textsc{ii}] $\lambda3727$ & $41.1\pm9.4$ & 1.793 & $59\pm13$\\
GN1 & 1823 & 189.131577 & 62.289875 & 24.2 & H$\alpha$ & $37.1\pm6.1$ & 0.537 & $55\pm9$\\
GN1 & 1831 & 189.159225 & 62.289642 & 25.0 & [O\textsc{iii}] $\lambda\lambda4959,5007$ & $87.2\pm10.2$ & 0.801 & $78\pm9$\\
GN1 & 1841 & 189.203278 & 62.28941 & 24.9 & [O\textsc{iii}] $\lambda\lambda4959,5007$ & $84.0\pm9.4$ & 0.955 & $104\pm11$\\
GN1 & 1957 & 189.180191 & 62.286591 & 25.1 & [O\textsc{iii}] $\lambda\lambda4959,5007$ & $120.3\pm10.8$ & 0.795 & $68\pm6$\\
GN1 & 1973 & 189.145004 & 62.286209 & 24.4 & [O\textsc{ii}] $\lambda3727$ & $34.9\pm7.8$ & 1.632 & $48\pm10$\\
GN1 & 2026 & 189.161438 & 62.285141 & 21.3 & H$\alpha$ & $379.0\pm65.1$ & 0.683 & $64\pm10$\\
GN1 & 2033 & 189.182953 & 62.284897 & 21.3 & H$\alpha$ & $186.4\pm34.4$ & 0.505 & $37\pm6$\\
GN1 & 2050 & 189.135483 & 62.283173 & 18.5 & H$\alpha$ & $54.4\pm27.1$ & 0.322 & $1\pm0$\\
GN1 & 2120 & 189.163193 & 62.28294 & 24.7 & [O\textsc{ii}] $\lambda3727$ & $16.0\pm5.1$ & 1.687 & $38\pm12$\\
GN1 & 2132 & 189.128632 & 62.282539 & 23.1 & [O\textsc{iii}] $\lambda\lambda4959,5007$ & $52.6\pm10.9$ & 0.94 & $52\pm11$\\
GN1 & 2135 & 189.162811 & 62.282536 & 23.6 & [O\textsc{iii}] $\lambda\lambda4959,5007$ & $150.4\pm12.3$ & 1.014 & $75\pm6$\\
GN1 & 2327 & 189.142426 & 62.278187 & 22.3 & H$\alpha$ & $105.2\pm20.7$ & 0.502 & $44\pm8$\\
GN1 & 2371 & 189.175491 & 62.277306 & 24.3 & [O\textsc{iii}] $\lambda\lambda4959,5007$ & $33.1\pm12.0$ & 0.949 & $60\pm22$\\
GN1 & 2394 & 189.164597 & 62.276897 & 25.8 & [O\textsc{iii}] $\lambda\lambda4959,5007$ & $28.5\pm10.4$ & 1.243 & $125\pm47$\\
GN1 & 2412 & 189.199005 & 62.276527 & 24.8 & [O\textsc{iii}] $\lambda\lambda4959,5007$ & $108.0\pm12.9$ & 0.779 & $76\pm9$\\
GN1 & 2449 & 189.181458 & 62.275795 & 23.3 & [O\textsc{ii}] $\lambda3727$ & $35.8\pm12.3$ & 1.487 & $37\pm13$\\
GN1 & 2713 & 189.188126 & 62.270935 & 23.5 & [O\textsc{ii}] $\lambda3727$ & $38.9\pm8.7$ & 1.445 & $53\pm11$\\
GN2 & 488 & 189.378052 & 62.325283 & 24.4 & [O\textsc{ii}] $\lambda3727$ & $34.6\pm9.0$ & 2.008 & $65\pm14$\\
GN2 & 506 & 189.35556 & 62.323696 & 23.4 & H$\alpha$ & $38.0\pm16.5$ & 0.335 & $26\pm10$\\
GN2 & 507 & 189.395798 & 62.323574 & 21.5 & H$\alpha$ & $153.7\pm29.4$ & 0.635 & $37\pm7$\\
GN2 & 514 & 189.370529 & 62.323437 & 24.3 & [O\textsc{iii}] $\lambda\lambda4959,5007$ & $64.3\pm8.1$ & 0.861 & $75\pm9$\\
GN2 & 554 & 189.397171 & 62.32164 & 20.7 & [O\textsc{iii}] $\lambda\lambda4959,5007$ & $285.0\pm33.6$ & 0.835 & $32\pm3$\\
GN2 & 591 & 189.34729 & 62.31987 & 24.9 & [O\textsc{ii}] $\lambda3727$ & $17.2\pm11.1$ & 1.995 & $61\pm34$\\
GN2 & 598 & 189.369843 & 62.319496 & 24.2 & [O\textsc{ii}] $\lambda3727$ & $27.5\pm7.1$ & 1.599 & $55\pm14$\\
GN2 & 657 & 189.358841 & 62.3158 & 23.6 & [O\textsc{ii}] $\lambda3727$ & $41.4\pm8.7$ & 1.595 & $49\pm10$\\
GN2 & 659 & 189.405548 & 62.315742 & 23.7 & [O\textsc{iii}] $\lambda\lambda4959,5007$ & $39.4\pm9.0$ & 1.08 & $63\pm13$\\
GN2 & 682 & 189.390213 & 62.319218 & 23.2 & [O\textsc{ii}] $\lambda3727$ & $66.9\pm18.0$ & 1.344 & $58\pm15$\\
GN2 & 717 & 189.414581 & 62.313183 & 22.4 & H$\alpha$ & $129.7\pm23.3$ & 0.338 & $46\pm8$\\
GN2 & 724 & 189.390701 & 62.312847 & 23.3 & [O\textsc{ii}] $\lambda3727$ & $73.4\pm10.9$ & 2.005 & $51\pm7$\\
GN2 & 740 & 189.349564 & 62.311909 & 23.0 & H$\alpha$ & $32.6\pm10.2$ & 0.558 & $38\pm12$\\
GN2 & 745 & 189.382751 & 62.311504 & 22.8 & [O\textsc{iii}] $\lambda\lambda4959,5007$ & $185.9\pm20.2$ & 1.084 & $98\pm10$\\
GN2 & 746 & 189.402069 & 62.311489 & 24.4 & [O\textsc{iii}] $\lambda\lambda4959,5007$ & $68.4\pm12.4$ & 1.086 & $113\pm174$\\
GN2 & 756 & 189.369843 & 62.310909 & 22.3 & H$\alpha$ & $81.7\pm12.5$ & 0.519 & $37\pm5$\\
GN2 & 757 & 189.421219 & 62.310848 & 22.1 & H$\alpha$ & $76.5\pm17.3$ & 0.576 & $38\pm8$\\
GN2 & 759 & 189.401138 & 62.310909 & 24.2 & [O\textsc{ii}] $\lambda3727$ & $32.8\pm8.4$ & 1.572 & $53\pm13$\\
GN2 & 780 & 189.347214 & 62.310238 & 24.4 & [O\textsc{iii}] $\lambda\lambda4959,5007$ & $58.7\pm9.4$ & 1.052 & $68\pm10$\\
GN2 & 782 & 189.387894 & 62.310043 & 24.5 & [O\textsc{iii}] $\lambda\lambda4959,5007$ & $33.6\pm7.2$ & 0.982 & $70\pm14$\\
GN2 & 814 & 189.33992 & 62.308514 & 24.2 & [O\textsc{ii}] $\lambda3727$ & $36.3\pm5.6$ & 1.973 & $72\pm10$\\
GN2 & 815 & 189.37941 & 62.308392 & 22.8 & [O\textsc{iii}] $\lambda\lambda4959,5007$ & $109.3\pm11.5$ & 1.197 & $67\pm7$\\
GN2 & 836 & 189.391006 & 62.307247 & 21.9 & H$\alpha$ & $101.1\pm19.3$ & 0.562 & $38\pm7$\\
GN2 & 852 & 189.398239 & 62.306988 & 24.2 & [O\textsc{ii}] $\lambda3727$ & $18.8\pm4.8$ & 2.051 & $42\pm10$\\
GN2 & 881 & 189.376816 & 62.305573 & 24.6 & [O\textsc{ii}] $\lambda3727$ & $31.9\pm6.5$ & 1.926 & $53\pm10$\\
GN2 & 909 & 189.416992 & 62.304211 & 25.5 & [O\textsc{ii}] $\lambda3727$ & $21.1\pm6.4$ & 1.781 & $85\pm26$\\
GN2 & 918 & 189.34906 & 62.303965 & 23.5 & [O\textsc{iii}] $\lambda\lambda4959,5007$ & $306.6\pm207.3$ & 1.078 & $121\pm93$\\
GN2 & 938 & 189.419174 & 62.302856 & 22.4 & [O\textsc{iii}] $\lambda\lambda4959,5007$ & $95.1\pm14.1$ & 1.027 & $47\pm7$\\
GN2 & 967 & 189.391983 & 62.301613 & 25.2 & [O\textsc{iii}] $\lambda\lambda4959,5007$ & $41.7\pm8.1$ & 1.224 & $71\pm13$\\
GN2 & 969 & 189.367142 & 62.30154 & 25.4 & [O\textsc{iii}] $\lambda\lambda4959,5007$ & $33.8\pm6.7$ & 1.146 & $75\pm14$\\
GN2 & 1049 & 189.385056 & 62.297539 & 24.2 & [O\textsc{ii}] $\lambda3727$ & $46.9\pm13.2$ & 2.006 & $71\pm17$\\
GN2 & 1065 & 189.364334 & 62.29715 & 23.9 & [O\textsc{iii}] $\lambda\lambda4959,5007$ & $77.1\pm12.2$ & 1.012 & $64\pm11$\\
GN2 & 1107 & 189.362579 & 62.295631 & 25.0 & [O\textsc{ii}] $\lambda3727$ & $14.2\pm8.4$ & 2.051 & $40\pm29$\\
GN2 & 1114 & 189.387207 & 62.29525 & 23.3 & [O\textsc{iii}] $\lambda\lambda4959,5007$ & $97.3\pm12.1$ & 0.773 & $54\pm6$\\
GN2 & 1145 & 189.355164 & 62.294254 & 24.1 & [O\textsc{iii}] $\lambda\lambda4959,5007$ & $31.8\pm9.9$ & 0.942 & $33\pm10$\\
GN2 & 1160 & 189.356705 & 62.293705 & 23.4 & [O\textsc{ii}] $\lambda3727$ & $36.3\pm8.9$ & 1.526 & $49\pm11$\\
GN2 & 1186 & 189.385986 & 62.292267 & 23.9 & [O\textsc{ii}] $\lambda3727$ & $31.3\pm6.1$ & 1.774 & $61\pm12$\\
GN2 & 1227 & 189.376297 & 62.290405 & 24.0 & [O\textsc{ii}] $\lambda3727$ & $36.2\pm10.6$ & 1.682 & $73\pm21$\\
GN2 & 1240 & 189.393906 & 62.289795 & 20.1 & H$\alpha$ & $194.1\pm18.9$ & 0.639 & $18\pm1$\\
GN2 & 1265 & 189.364151 & 62.289097 & 24.6 & H$\alpha$ & $27.4\pm4.7$ & 0.633 & $50\pm8$\\
GN2 & 1319 & 189.387177 & 62.287018 & 22.7 & H$\alpha$ & $90.1\pm18.5$ & 0.632 & $63\pm13$\\
GN2 & 3114 & 189.376511 & 62.325085 & 25.5 & [O\textsc{iii}] $\lambda\lambda4959,5007$ & $31.8\pm11.6$ & 0.839 & $135\pm91$\\
GN2 & 3574 & 189.389481 & 62.315483 & 27.1 & [O\textsc{iii}] $\lambda\lambda4959,5007$ & $54.2\pm10.1$ & 0.863 & $57\pm11$\\
GN2 & 4969 & 189.363007 & 62.287544 & 25.9 & [O\textsc{ii}] $\lambda3727$ & $47.7\pm15.3$ & 1.363 & $423\pm171$\\
GS1 & 724 & 53.172264 & -27.760622 & 22.6 & [O\textsc{ii}] $\lambda3727$ & $68.7\pm19.8$ & 1.552 & $39\pm11$\\
GS1 & 950 & 53.161499 & -27.76762 & 23.4 & H$\alpha$ & $38.1\pm10.5$ & 0.679 & $34\pm7$\\
GS1 & 970 & 53.160183 & -27.769306 & 24.1 & [O\textsc{iii}] $\lambda\lambda4959,5007$ & $69.0\pm11.2$ & 1.044 & $88\pm14$\\
GS1 & 1013 & 53.169926 & -27.771027 & 20.0 & H$\alpha$ & $156.7\pm25.6$ & 0.618 & $14\pm2$\\
GS1 & 1016 & 53.172104 & -27.770382 & 23.6 & H$\alpha$ & $18.2\pm3.8$ & 0.631 & $34\pm7$\\
GS1 & 1056 & 53.162453 & -27.770908 & 24.3 & [O\textsc{iii}] $\lambda\lambda4959,5007$ & $35.6\pm4.2$ & 1.036 & $88\pm10$\\
GS1 & 1103 & 53.174 & -27.772057 & 20.7 & H$\alpha$ & $198.7\pm27.6$ & 0.335 & $23\pm3$\\
GS1 & 1132 & 53.184479 & -27.772245 & 24.6 & [O\textsc{iii}] $\lambda\lambda4959,5007$ & $31.9\pm7.6$ & 0.835 & $50\pm12$\\
GS1 & 1151 & 53.152878 & -27.772497 & 24.6 & [O\textsc{ii}] $\lambda3727$ & $20.0\pm7.4$ & 1.851 & $48\pm18$\\
GS1 & 1171 & 53.151215 & -27.772837 & 23.8 & H$\alpha$ & $49.4\pm8.2$ & 0.607 & $47\pm7$\\
GS1 & 1239 & 53.191463 & -27.77389 & 23.5 & H$\alpha$ & $32.7\pm9.7$ & 0.419 & $36\pm10$\\
GS1 & 1295 & 53.162361 & -27.775063 & 20.6 & H$\alpha$ & $352.9\pm61.4$ & 0.419 & $39\pm6$\\
GS1 & 1296 & 53.159355 & -27.775028 & 23.1 & [O\textsc{iii}] $\lambda\lambda4959,5007$ & $41.9\pm16.9$ & 1.22 & $45\pm18$\\
GS1 & 1299 & 53.160801 & -27.775373 & 21.2 & H$\alpha$ & $110.2\pm27.4$ & 0.623 & $27\pm6$\\
GS1 & 1359 & 53.185909 & -27.775608 & 22.9 & [O\textsc{ii}] $\lambda3727$ & $78.3\pm20.6$ & 1.425 & $60\pm15$\\
GS1 & 1392 & 53.181046 & -27.776175 & 22.1 & H$\alpha$ & $102.0\pm22.9$ & 0.668 & $40\pm8$\\
GS1 & 1467 & 53.151047 & -27.777309 & 24.1 & [O\textsc{iii}] $\lambda\lambda4959,5007$ & $32.2\pm7.6$ & 0.733 & $55\pm12$\\
GS1 & 1467 & 53.151047 & -27.777309 & 24.1 & H$\alpha$ & $28.4\pm10.6$ & 0.733 & $52\pm21$\\
GS1 & 1476 & 53.147438 & -27.777596 & 23.6 & [O\textsc{ii}] $\lambda3727$ & $30.5\pm8.5$ & 1.851 & $48\pm13$\\
GS1 & 1477 & 53.158291 & -27.777449 & 24.5 & [O\textsc{ii}] $\lambda3727$ & $26.7\pm6.6$ & 1.557 & $41\pm10$\\
GS1 & 1481 & 53.146614 & -27.777489 & 25.0 & [O\textsc{iii}] $\lambda\lambda4959,5007$ & $20.5\pm12.3$ & 1.091 & $47\pm24$\\
GS1 & 1500 & 53.152336 & -27.777948 & 24.3 & [O\textsc{ii}] $\lambda3727$ & $30.0\pm7.1$ & 1.42 & $50\pm12$\\
GS1 & 1552 & 53.157204 & -27.778522 & 23.8 & [O\textsc{ii}] $\lambda3727$ & $24.2\pm22.0$ & 1.31 & $34\pm29$\\
GS1 & 1689 & 53.162483 & -27.780346 & 25.1 & H$\alpha$ & $37.9\pm8.7$ & 0.722 & $52\pm12$\\
GS1 & 1689 & 53.162483 & -27.780346 & 25.1 & [O\textsc{iii}] $\lambda\lambda4959,5007$ & $63.6\pm12.3$ & 0.722 & $65\pm12$\\
GS1 & 1710 & 53.172619 & -27.78096 & 21.4 & H$\alpha$ & $111.4\pm24.5$ & 0.62 & $32\pm7$\\
GS1 & 1711 & 53.196999 & -27.780598 & 23.9 & [O\textsc{iii}] $\lambda\lambda4959,5007$ & $36.0\pm7.2$ & 0.739 & $46\pm8$\\
GS1 & 1728 & 53.176331 & -27.780861 & 25.0 & [O\textsc{iii}] $\lambda\lambda4959,5007$ & $42.0\pm10.1$ & 1.016 & $60\pm14$\\
GS1 & 1803 & 53.170067 & -27.782066 & 26.7 & [O\textsc{ii}] $\lambda3727$ & $55.1\pm8.8$ & 1.344 & $81\pm12$\\
GS1 & 1829 & 53.150764 & -27.78256 & 24.2 & [O\textsc{ii}] $\lambda3727$ & $74.6\pm9.7$ & 1.343 & $92\pm11$\\
GS1 & 1851 & 53.152782 & -27.782698 & 24.4 & [O\textsc{iii}] $\lambda\lambda4959,5007$ & $101.1\pm10.5$ & 0.768 & $72\pm7$\\
GS1 & 1864 & 53.175331 & -27.782722 & 26.0 & [O\textsc{iii}] $\lambda\lambda4959,5007$ & $36.9\pm19.6$ & 0.843 & $152\pm81$\\
GS1 & 1867 & 53.15184 & -27.782864 & 23.2 & H$\alpha$ & $66.3\pm13.6$ & 0.406 & $40\pm8$\\
GS1 & 1900 & 53.184574 & -27.783323 & 24.4 & [O\textsc{iii}] $\lambda\lambda4959,5007$ & $98.8\pm15.8$ & 1.137 & $76\pm12$\\
GS1 & 1946 & 53.192593 & -27.783791 & 24.7 & [O\textsc{iii}] $\lambda\lambda4959,5007$ & $65.8\pm11.7$ & 0.869 & $105\pm18$\\
GS1 & 2023 & 53.151863 & -27.784752 & 25.6 & [O\textsc{iii}] $\lambda\lambda4959,5007$ & $33.0\pm9.6$ & 1.217 & $86\pm25$\\
GS1 & 2029 & 53.157948 & -27.784767 & 28.1 & [O\textsc{iii}] $\lambda\lambda4959,5007$ & $22.4\pm9.5$ & 0.719 & $34\pm14$\\
GS1 & 2029 & 53.157948 & -27.784767 & 28.1 & H$\alpha$ & $15.1\pm7.7$ & 0.719 & $23\pm11$\\
GS1 & 2039 & 53.166565 & -27.784861 & 27.6 & [O\textsc{ii}] $\lambda3727$ & $17.2\pm9.0$ & 1.304 & $109\pm57$\\
GS1 & 2077 & 53.161686 & -27.785322 & 27.5 & H$\alpha$ & $29.7\pm9.4$ & 0.337 & $87\pm27$\\
GS1 & 2138 & 53.160477 & -27.786299 & 24.3 & [O\textsc{iii}] $\lambda\lambda4959,5007$ & $35.3\pm7.5$ & 0.983 & $57\pm11$\\
GS1 & 2168 & 53.163471 & -27.786636 & 25.1 & H$\alpha$ & $17.6\pm5.1$ & 0.469 & $47\pm14$\\
GS1 & 2187 & 53.177753 & -27.786966 & 24.4 & [O\textsc{iii}] $\lambda\lambda4959,5007$ & $38.3\pm9.9$ & 0.95 & $75\pm20$\\
GS1 & 2221 & 53.164097 & -27.787298 & 23.8 & [O\textsc{iii}] $\lambda\lambda4959,5007$ & $190.9\pm12.2$ & 1.098 & $64\pm4$\\
GS1 & 2291 & 53.149296 & -27.788527 & 23.0 & [O\textsc{ii}] $\lambda3727$ & $65.3\pm15.1$ & 1.916 & $52\pm12$\\
GS1 & 2338 & 53.15736 & -27.789219 & 25.1 & [O\textsc{iii}] $\lambda\lambda4959,5007$ & $35.5\pm9.4$ & 0.999 & $85\pm24$\\
GS1 & 2363 & 53.168015 & -27.789671 & 22.8 & H$\alpha$ & $79.4\pm16.8$ & 0.621 & $55\pm11$\\
GS1 & 2375 & 53.176495 & -27.789705 & 24.9 & H$\alpha$ & $26.1\pm8.1$ & 0.427 & $42\pm13$\\
GS1 & 2378 & 53.18795 & -27.790001 & 20.3 & H$\alpha$ & $749.5\pm279.4$ & 0.438 & $47\pm17$\\
GS1 & 2385 & 53.184811 & -27.789934 & 23.1 & [O\textsc{iii}] $\lambda\lambda4959,5007$ & $43.2\pm10.7$ & 0.956 & $45\pm11$\\
GS1 & 2417 & 53.160419 & -27.790369 & 23.5 & [O\textsc{ii}] $\lambda3727$ & $41.5\pm8.7$ & 1.617 & $44\pm9$\\
GS1 & 2495 & 53.184132 & -27.791531 & 23.1 & [O\textsc{iii}] $\lambda\lambda4959,5007$ & $7.3\pm10.7$ & 1.214 & $8\pm11$\\
GS1 & 2517 & 53.161613 & -27.792299 & 20.6 & H$\alpha$ & $1404.0\pm251.4$ & 0.462 & $65\pm11$\\
GS1 & 2560 & 53.184158 & -27.792637 & 21.4 & [O\textsc{iii}] $\lambda\lambda4959,5007$ & $162.6\pm18.2$ & 0.739 & $107\pm12$\\
GS1 & 2560 & 53.184158 & -27.792637 & 21.4 & H$\alpha$ & $203.3\pm53.7$ & 0.739 & $84\pm23$\\
GS1 & 2570 & 53.164124 & -27.792654 & 27.0 & [O\textsc{ii}] $\lambda3727$ & $16.5\pm31.8$ & 1.301 & $36\pm64$\\
GS1 & 2654 & 53.182205 & -27.793993 & 24.8 & H$\alpha$ & $20.8\pm11.7$ & 1.28 & $60\pm33$\\
GS1 & 2669 & 53.156631 & -27.794302 & 24.3 & [O\textsc{iii}] $\lambda\lambda4959,5007$ & $41.5\pm10.3$ & 1.098 & $61\pm14$\\
GS1 & 2696 & 53.155861 & -27.794901 & 22.9 & [O\textsc{iii}] $\lambda\lambda4959,5007$ & $104.1\pm31.9$ & 1.1 & $74\pm22$\\
GS1 & 2720 & 53.15675 & -27.79558 & 21.8 & [O\textsc{iii}] $\lambda\lambda4959,5007$ & $90.5\pm18.7$ & 1.099 & $42\pm11$\\
GS1 & 2732 & 53.161331 & -27.795797 & 23.6 & [O\textsc{ii}] $\lambda3727$ & $50.7\pm13.2$ & 1.495 & $50\pm13$\\
GS1 & 2783 & 53.188084 & -27.795742 & 24.0 & H$\alpha$ & $116.4\pm8.9$ & 0.536 & $47\pm3$\\
GS1 & 2872 & 53.16687 & -27.797707 & 23.7 & [O\textsc{iii}] $\lambda\lambda4959,5007$ & $70.2\pm14.2$ & 0.99 & $75\pm15$\\
GS1 & 2942 & 53.161121 & -27.798801 & 25.5 & [O\textsc{iii}] $\lambda\lambda4959,5007$ & $27.9\pm13.5$ & 1.228 & $82\pm34$\\
GS1 & 4184 & 53.179535 & -27.766174 & 25.1 & H$\alpha$ & $15.7\pm5.8$ & 0.67 & $35\pm12$\\
GS1 & 4198 & 53.178375 & -27.76824 & 20.2 & H$\alpha$ & $341.1\pm49.8$ & 0.674 & $32\pm4$\\
GS1 & 4258 & 53.152287 & -27.770088 & 23.7 & [O\textsc{ii}] $\lambda3727$ & $45.4\pm14.5$ & 1.854 & $50\pm16$\\
GS1 & 4284 & 53.184544 & -27.768221 & 25.2 & [O\textsc{ii}] $\lambda3727$ & $7.2\pm4.3$ & 1.842 & $14\pm8$\\
GS1 & 6865 & 53.190331 & -27.774298 & 26.8 & [O\textsc{iii}] $\lambda\lambda4959,5007$ & $18.2\pm12.1$ & 0.883 & $54\pm35$\\
GS1 & 8178 & 53.187664 & -27.783779 & 27.0 & [O\textsc{iii}] $\lambda\lambda4959,5007$ & $22.0\pm7.6$ & 0.737 & $110\pm38$\\
GS2 & 575 & 53.28241 & -27.843513 & 24.0 & [O\textsc{iii}] $\lambda\lambda4959,5007$ & $87.8\pm16.6$ & 0.739 & $73\pm14$\\
GS2 & 575 & 53.28241 & -27.843513 & 24.0 & H$\alpha$ & $68.5\pm16.9$ & 0.739 & $61\pm15$\\
GS2 & 577 & 53.273159 & -27.844625 & 25.9 & [O\textsc{iii}] $\lambda\lambda4959,5007$ & $24.2\pm9.7$ & 1.231 & $66\pm25$\\
GS2 & 596 & 53.274158 & -27.84565 & 23.2 & [O\textsc{ii}] $\lambda3727$ & $34.8\pm7.4$ & 1.686 & $36\pm7$\\
GS2 & 599 & 53.279076 & -27.845737 & 23.1 & [O\textsc{iii}] $\lambda\lambda4959,5007$ & $318.9\pm40.1$ & 0.737 & $71\pm8$\\
GS2 & 599 & 53.279076 & -27.845737 & 23.1 & H$\alpha$ & $228.1\pm59.8$ & 0.737 & $55\pm14$\\
GS2 & 620 & 53.272892 & -27.847765 & 22.2 & H$\alpha$ & $81.0\pm9.1$ & 0.711 & $35\pm3$\\
GS2 & 709 & 53.288483 & -27.851877 & 24.0 & H$\alpha$ & $22.4\pm5.3$ & 0.687 & $46\pm11$\\
GS2 & 782 & 53.281536 & -27.854385 & 23.7 & [O\textsc{iii}] $\lambda\lambda4959,5007$ & $81.2\pm10.8$ & 0.834 & $56\pm7$\\
GS2 & 846 & 53.264668 & -27.855431 & 24.6 & [O\textsc{ii}] $\lambda3727$ & $26.7\pm9.7$ & 1.76 & $56\pm20$\\
GS2 & 868 & 53.275829 & -27.855747 & 24.7 & H$\alpha$ & $33.4\pm9.6$ & 0.737 & $82\pm23$\\
GS2 & 871 & 53.266712 & -27.856167 & 20.5 & H$\alpha$ & $559.4\pm184.3$ & 0.529 & $49\pm16$\\
GS2 & 887 & 53.291748 & -27.856255 & 24.2 & [O\textsc{ii}] $\lambda3727$ & $28.5\pm11.2$ & 1.815 & $56\pm21$\\
GS2 & 951 & 53.264828 & -27.857828 & 24.3 & [O\textsc{ii}] $\lambda3727$ & $66.2\pm17.7$ & 1.303 & $68\pm17$\\
GS2 & 951 & 53.264828 & -27.857828 & 24.3 & [O\textsc{iii}] $\lambda\lambda4959,5007$ & $235.2\pm20.1$ & 1.291 & $107\pm9$\\
GS2 & 1038 & 53.285847 & -27.85964 & 20.6 & [O\textsc{iii}] $\lambda\lambda4959,5007$ & $238.1\pm97.6$ & 0.725 & $33\pm13$\\
GS2 & 1038 & 53.285847 & -27.85964 & 20.6 & H$\alpha$ & $754.1\pm49.5$ & 0.725 & $66\pm4$\\
GS2 & 1054 & 53.2869 & -27.859509 & 22.8 & [O\textsc{ii}] $\lambda3727$ & $56.6\pm19.4$ & 1.68 & $37\pm12$\\
GS2 & 1131 & 53.26556 & -27.861135 & 23.6 & [O\textsc{iii}] $\lambda\lambda4959,5007$ & $37.9\pm14.6$ & 0.885 & $44\pm14$\\
GS2 & 1215 & 53.266247 & -27.862015 & 23.9 & [O\textsc{ii}] $\lambda3727$ & $21.3\pm8.4$ & 1.905 & $37\pm14$\\
GS2 & 1240 & 53.293732 & -27.862436 & inf & [O\textsc{ii}] $\lambda3727$ & $14.8\pm3.2$ & 1.901 & $49\pm11$\\
GS2 & 1270 & 53.275627 & -27.863014 & 24.4 & [O\textsc{ii}] $\lambda3727$ & $29.4\pm7.8$ & 2.012 & $72\pm19$\\
GS2 & 1280 & 53.283585 & -27.864466 & 22.9 & H$\alpha$ & $74.4\pm12.2$ & 0.476 & $43\pm6$\\
GS2 & 1392 & 53.28775 & -27.865278 & 24.3 & H$\alpha$ & $22.2\pm5.2$ & 0.612 & $44\pm10$\\
GS2 & 1483 & 53.272259 & -27.867044 & 24.7 & [O\textsc{iii}] $\lambda\lambda4959,5007$ & $61.9\pm8.5$ & 0.979 & $73\pm10$\\
GS2 & 1552 & 53.266247 & -27.868002 & 22.9 & [O\textsc{ii}] $\lambda3727$ & $81.0\pm21.8$ & 1.761 & $50\pm13$\\
GS2 & 1593 & 53.272778 & -27.868891 & 21.6 & H$\alpha$ & $135.9\pm37.2$ & 0.693 & $39\pm10$\\
GS2 & 1607 & 53.265091 & -27.86924 & 21.7 & H$\alpha$ & $126.0\pm32.7$ & 0.524 & $46\pm12$\\
GS2 & 1630 & 53.264194 & -27.869268 & 23.3 & [O\textsc{ii}] $\lambda3727$ & $53.7\pm18.7$ & 1.817 & $51\pm18$\\
GS2 & 1653 & 53.273643 & -27.870647 & 18.4 & H$\alpha$ & $515.4\pm85.3$ & 0.524 & $12\pm2$\\
GS2 & 1666 & 53.268139 & -27.869875 & 24.2 & [O\textsc{iii}] $\lambda\lambda4959,5007$ & $97.4\pm12.3$ & 0.737 & $80\pm10$\\
GS2 & 1666 & 53.268139 & -27.869875 & 24.2 & H$\alpha$ & $52.5\pm10.7$ & 0.737 & $53\pm10$\\
GS2 & 1772 & 53.28323 & -27.872059 & 25.8 & [O\textsc{iii}] $\lambda\lambda4959,5007$ & $53.4\pm9.5$ & 1.176 & $73\pm12$\\
GS2 & 1836 & 53.2654 & -27.873278 & 22.8 & [O\textsc{ii}] $\lambda3727$ & $105.2\pm22.1$ & 1.988 & $57\pm12$\\
GS2 & 1845 & 53.276325 & -27.873322 & 26.1 & [O\textsc{iii}] $\lambda\lambda4959,5007$ & $18082.0\pm2131.0$ & 1.044 & $51\pm6$\\
GS2 & 3186 & 53.291828 & -27.845343 & inf & H$\alpha$ & $1500.3\pm231.8$ & 0.523 & $63\pm9$\\
GS2 & 3259 & 53.276016 & -27.847622 & 24.4 & [O\textsc{iii}] $\lambda\lambda4959,5007$ & $339.3\pm25.4$ & 1.262 & $99\pm10$\\
GS2 & 3277 & 53.277046 & -27.848417 & 24.6 & [O\textsc{ii}] $\lambda3727$ & $21.0\pm4.0$ & 1.907 & $38\pm7$\\
GS2 & 3295 & 53.258167 & -27.849049 & 23.4 & [O\textsc{ii}] $\lambda3727$ & $96.4\pm23.2$ & 1.9 & $65\pm15$\\
GS2 & 3314 & 53.285191 & -27.850237 & 24.3 & [O\textsc{ii}] $\lambda3727$ & $27.5\pm8.5$ & 1.717 & $63\pm19$\\
GS2 & 3347 & 53.266071 & -27.852331 & 22.0 & H$\alpha$ & $120.6\pm26.7$ & 0.548 & $50\pm11$\\
GS2 & 3418 & 53.253826 & -27.856579 & 27.0 & [O\textsc{iii}] $\lambda\lambda4959,5007$ & $39.6\pm12.4$ & 0.909 & $92\pm28$\\
GS2 & 3419 & 53.254498 & -27.856409 & 24.3 & H$\alpha$ & $26.7\pm5.4$ & 0.521 & $42\pm8$\\
\enddata
\end{deluxetable*}



\end{document}